%% file: ms_uchii_maser_WK.tex
\documentclass[preprint]{aastex6}

\usepackage{graphicx}
\usepackage{longtable}
\usepackage{float}
\usepackage{natbib}
\usepackage{soul}
\usepackage[normalem]{ulem}
\usepackage[colorinlistoftodos,prependcaption,textsize=tiny]{todonotes}

\newcommand{\ghz}{\mbox{GHz}}
\newcommand{\kms}{\mbox{km\,s$^{-1}$}}

\newcommand{\cm}{\mbox{cm}}

\newcommand{\methanol}{\mbox{CH$_3$OH}}

\newcommand{\water}{\mbox{H$_2$O}}
\newcommand{\ta}{\mbox{$T_{\rm A}^*$}}
\newcommand{\hii}{\mbox{H{\sc ii}}}

\newcommand{\uchii}{\mbox{UCH{\sc ii}}}

\newcommand{\uchiis}{\mbox{UCH{\sc ii}s}}
\newcommand{\msun}{M$_\odot$}

\usepackage{xcolor}
\newif\ifmod
\modtrue
\newif\ifdel
\delfalse

\begin{document}


\title{Simultaneous 22\,GHz Water and 44\,GHz Methanol Masers Survey of Ultracompact \hii\ Regions}



\author{Won-Ju Kim$^{1,2,3}$, 
		\and Kee-Tae Kim$^{1, 4}$, 
        and Kwang-Tae Kim$^{2}$} 
\affil{
$^1$ Korea Astronomy and Space Science Institute, 776 Daedeokdae-ro, Yuseong-gu, Daejeon 34055, Republic of Korea \\
$^2$ Chungnam National University, Yuseong-gu, Daejeon 34134, Republic of Korea \\
$^3$ Instituto de Radioastronom\'ia Milim\'etrica, Avenida Divina Pastora 7, 18012 Granada, Spain \\
$^4$ University of Science and Technology, Korea (UST), 217 Gajeong-ro, Yuseong-gu, Daejeon 34113, Republic of Korea}



\begin{abstract}
We have carried out a multi-epoch simultaneous survey of 22\,\ghz\ \water\ and 44\,\ghz\ Class\,I \methanol\ masers toward 103 ultracompact \hii\ regions (\uchiis) between 2010 and 2017. We detected \water\ and \methanol\  maser emission in 74 (72\%) and 55 (53\%) \uchiis, respectively. Among them, 3 \water\ and 27 \methanol\ maser sources are new detections. These high detection rates suggest that the occurrence periods of both maser species are significantly overlapped with the \uchii\ phase of massive star formation. The \methanol\ maser lines always have small ($<$\,10\,\kms) relative velocities with respect to the natal molecular cores, while \water\ maser lines often show larger relative velocities. Twenty four \water\ maser-detected sources have maser lines at relative velocities $>$\,30\,\kms, and thirteen of them show extremely high-velocity ($>$\,50\,\kms) components. The appearance and disappearance of \water\ maser lines are quite frequent over six-month or one-year time intervals. In contrast, \methanol\ maser lines usually do not exhibit significant variations in the line intensity, velocity, or shape for the same periods. The isotropic luminosities of both masers appear to correlate with the bolometric luminosities of the central stars. This correlation becomes much stronger in the case that data points in the low- and intermediate-mass regimes are added. The maser luminosities also tend to increase with the radio continuum luminosities of \uchiis\ and the submillimeter continuum luminosities of the associated dense cores. 
\end{abstract}

\maketitle

\section{INTRODUCTION}\label{sec:intro}
Various maser species are detected in star-forming regions and used to understand the fundamental processes of star formation, such as accretion disks, jets/outflows, and inflows. Water (\water), methanol (\methanol), and hydroxyl (OH) masers are the major maser species associated with star-forming regions. \water\ and OH masers are observed both in star-forming regions and evolved stars, while \methanol\ masers are exclusively detected in star-forming regions, especially high-mass ($>$ 8\,\msun) star-forming regions (e.g., \citealt{garay1999,breen2013,urquhart2015}). \water\ masers at 22\,\ghz\ frequency have been extensively studied since the first discovery by \cite{cheung1969} in the interstellar medium. They are detected in a number of star-forming regions. In low- and intermediate-mass star-forming regions, \water\ masers are mostly associated with protostars (Class 0/I objects), while they are rarely detected in pre-main-sequence stars \citep{furuya2003,bae2011}. In high-mass star-forming regions, on the other hand, the masers are detected in a wide range of evolutionary stages, from protostellar objects to ultracompact \hii\ regions (\uchiis), the central stars of which have already reached the main-sequence stage (e.g., \citealt{churchwell1990,hofner1996,beuther2002,urquhart2009,urquhart2011,kim2018a}). \cite{elitzur1989} suggested, through detailed numerical calculations, that \water\ masers could be generated by collisional pumping in the post-shocked regions with unusually high gas temperature and density ($T_{\rm kin}\,\sim\,$500\,K and $n_{\rm H_2}\,\sim\,10^{9}$\,cm$^{-3}$). \water\ masers were observed to be closely related to outflows \citep{felli1992}. High-resolution observations show that they are distributed in the immediate vicinity of the central stars and usually trace inner jets/outflows (e.g., \citealt{torrelles1997,torrelles2011,goddi2006}). The \water\ maser luminosity has been found to increase with the bolometric luminosity of the central star \citep{wouterloot1986, palla1991, palla1993, felli1992, wilking1994, furuya2003, bae2011}.

There are more than 40 \methanol\ molecular transitions, which have been observed to show maser emission \citep{muller2004}. The \methanol\ maser transitions were divided into two groups, Class\,I and Class\,II, based on their association with other star formation indicators, such as IRAS point sources, \uchiis, high-mass protostellar objects (HMPOs), hot molecular cores (HMCs), \water\ maser sources, and so on \citep{menten1991}. Class\,II \methanol\ masers are more spatially correlated with other star formation indicators, while Class\,I \methanol\ masers tend to have offsets from those indicators. \cite{kurtz2004} found offsets of 0.1$-$1\,pc between 44\,\ghz\ Class\,I \methanol\ masers and \uchiis\ in their subarcsecond-resolution observations using the Very Large Array (VLA). According to theoretical studies \citep{menten1996,sobolev2007,cragg2005}, Class\,I \methanol\ masers can be produced by collisional pumping in regions with $T_{\rm kin}\sim\,$100\,K and $n_{\rm H_2}\sim\,$10$^{5}$\,cm$^{-3}$, whereas Class\,II \methanol\ masers can be excited by radiation pumping in hotter and denser regions ($T_{\rm kin}\sim\,$150\,K and $n_{\rm H_2}\sim\,$10$^{7}$\,cm$^{-3}$). These results suggest that the emitting regions of the two classes of maser emission are different. Interferometric observations show that Class\,II masers originate from accretion disks and inner jets (e.g., \citealt{bartkiewicz2009, bartkiewicz2011}), while Class\,I masers emanate from the interacting regions of jets/outflows with the ambient dense molecular gas (e.g., \citealt{plambeck1990, kurtz2004}). The most well-known \methanol\ maser lines in Class\,I include 44\,\ghz\ ($7_{0}-6_{1}\,A^{+}$) and 95\,\ghz\ ($8_{0}-7_{1}~A^{+}$) transitions and in  Class\,II contain 6.7\,\ghz\ ($5_{1}-6_{0}\,A^{+}$) and 12.2\,\ghz\ ($2_{0}-3_{1}\,E$) transitions.

\uchiis\ are very small and dense hydrogen ionized gas regions with diameters $\lesssim$\,0.1\,pc, electron densities ($n_{e}$) $\gtrsim$\,10$^4$\,cm$^{-3}$ and emission measures (EM) $\gtrsim$\,10$^7$\,pc\,cm$^{-6}$ (\citealt{wood1989a}, hereafter WC89a; \citealt{kurtz1994}, hereafter KCW94). They are not only strong radio sources but also bright infrared sources, suggesting that they are still deeply embedded in the parental molecular clouds. Since they can be produced only by O- or early B-type stars, they have been believed to be prominent signposts of massive star-forming sites. In fact, \cite{shepherd1996} showed that many \uchiis\ are associated with bipolar molecular outflows, which are salient indicators of star formation. The vast majority of them are also associated with masers of \water, \methanol, and/or OH (see \citealt{garay1999} for a review).

In this paper, we present a multi-epoch simultaneous survey of 22\,\ghz\ \water\ and 44\,\ghz\ Class\,I \methanol\ masers toward 103 \uchiis. Particularly, the present survey provides a new 44~\ghz\ \methanol\ maser catalog covering a large sample of \uchiis, with much higher sensitivities than most previous single-dish surveys. The primary scientific goal of this survey is to investigate the characteristics of the two maser species associated with \uchiis\ and to examine the relationship between these maser species and the evolutionary sequence of massive star formation. The source selection and observational details are described in \S\,\ref{sec:source} and \S\, \ref{sec:maser_obs}, respectively. We present the results in \S\,\ref{sec:result} and comments on some interesting individual sources in \S\,\ref{sec:individual}. In \S\,\ref{sec:discussion}, we compare properties of the two masers with those of the central stars and host clumps and discuss the implications for the evolutionary sequence of massive star formation. The main results are summarized in \S\,\ref{sec:summary}.

\section{OBSERVATIONS}\label{sec:obs}

\subsection{Source Selection}\label{sec:source}
We selected 103 \uchiis\ from the catalogs of WC89a and KCW94. WC89a performed subarcsecond-resolution radio continuum observations with the VLA toward 80 compact \hii\ regions and complexes at 2 and 6\,cm wavelengths. They detected at least one \uchii\ in each of 53 regions. \cite{wood1989b} found that all known \uchiis\ at that time have characteristic IRAS colors, Log$(F_{\rm 60}/F_{\rm 12})\,\ge\,1.30$ and Log$(F_{\rm 25}/F_{\rm 12})\,\ge\,0.57$, and then identified $\sim$\,1650 \uchii\ candidates from the IRAS point source catalog based on these color criteria. KCW94 made VLA radio continuum observations of 59 \uchii\ candidates with $F_{\rm 100}\,\ge$\,1000\,Jy and the \hii\ region complex of DR21, to validate the IRAS two-color selection criteria. They detected radio continuum emission toward 49 (82\%) of the observed sources. Of the remaining 11 sources, IRAS05391$-$0152 and IRAS19410$+$2336 have been confirmed as \uchiis\ afterward \citep{walsh1998, carral1999}. There is one common source between the two catalogs, G35.20$-$1.74 in WC89a and IRAS18592+0108 in KCW94. In summary, the sample of this study consists of 52 \uchiis\ from WC89a and 51 \uchiis\ from KCW94 (see Table\,\ref{tb:source}).

\subsection{22 \ghz\ \water\ and the 44 \ghz\ \methanol\ Maser Observations}\label{sec:maser_obs}

We carried out a four-epoch simultaneous \water\ (6$_{1,6}-$5$_{2,3}$, $\nu$ = 22.23508\,\ghz) and Class\,I \methanol\ (7$_{0}-$6$_{1}$ A$^{+}$, $\nu$ = 44.06943\,\ghz) maser survey toward the \uchiis\ in our sample between 2010 and 2017 using the Korean VLBI Network (KVN) 21 m radio telescopes \citep{kim2011,lee2011}. The telescopes were equipped with the multi-frequency receiving systems, which makes it possible to observe at 22 and 44\,\ghz\ frequencies simultaneously \citep{han2008}. The 4096-channel digital spectrometers provide 32\,MHz bandwidth corresponding velocity coverage of 432 and 218\,\kms\ at 22 and 44\,\ghz, respectively. The pointing and focus of the telescope were checked every 2$-$3 hours by observing strong  \water\ and 43\,\ghz\ SiO ($v$=1, J=1$-$0) maser sources, such as V1111~Oph, R~Cas, W3(OH), and Orion~KL, as calibrators. The pointing accuracy was better than 5\arcsec. All spectra were obtained in the position switching mode, and the total ON+OFF integration time per source was usually 30 minutes yielding a typical $rms$ noise level of $\sim$\,0.5\,Jy at a velocity resolution of 0.21\,\kms\ after smoothing once at 22~\ghz\ and twice at 44~\ghz. The data were calibrated using the standard chopper wheel method, and the line intensity was obtained on the \ta\ scale. The conversion factors of \ta\ to flux density are 11.07 and 11.60\,Jy\,K$^{-1}$ at 22 and 44\,GHz, respectively, assuming that the aperture efficiencies ($\eta_{a}$) of the telescope are 0.72 and 0.69 at the observing frequencies \citep{lee2011}. The full-width half maxima (FWHMs) of the main beams are 130\arcsec\ at 22\,\ghz\ and 65\arcsec\ at 44\,\ghz. The data reduction was performed using CLASS software of GILDAS package.

The first- and second-epoch observations were conducted toward all the \uchiis\ in our sample, except G43.18$-$0.52, in 2010 and 2011, respectively. The observed positions were mostly the radio continuum emission peaks for the WC89a subsample (their Table~4) and the associated IRAS point source positions for the KCW94 subsample (their Table~2). Afterward we found that the observed positions were offset by larger than 10$''$ from the radio continuum peaks for 43 sources (Tables 11$-$14 of WC89a and Table~3 of KCW94): $10''-20''$ for 21 sources, $20''-30''$ for 6 sources, $>30''$ for 16 sources (see the 5th column of Table\,\ref{tb:source}). The third-epoch observations were undertaken toward the radio continuum emission peaks of those 43 \uchiis\ and G43.18$-$0.52 in 2017 June, and the fourth-epoch observations were made toward only 28 out of the 44 \uchiis, including all the 22 sources with offsets $>20''$, in 2017 June. For the remaining 59 sources with offsets $<10''$, the peak fluxes could be reduced due to the positional uncertainty by up to 0.4$-$3.6\% and 1.6$-$13.7\% at 22 and 44~\ghz, respectively, assuming that the main-beam patterns are Gaussian with the aforementioned FWHMs.

Since the KVN telescopes are shaped Cassegrain type, they have much higher first sidelobe levels than the conventional Cassegrain telescopes, $\sim$\,14\,dB versus 20$-$30\,dB \citep{kim2011,lee2011}. The first sidelobe is about 1.5\,$\times$\,FWHM away from the main beam center. We checked whether the detected maser emission was contaminated by nearby strong sources, by mapping an area of 3.0\,FWHM\,$\times$\,3.0\,FWHM around each maser-detected source at half-beam grid spacing. In the first and second epochs, maser emission was found to be detected by the first sidelobe in 6 sources: IRAS19081$+$0903, G75.83$+$0.4, and DR21 at 22\,\ghz, and IRAS06099$+$1800, G10.62$-$0.38, and G23.46$-$0.20 at 44\,\ghz. In the third and fourth epochs, however, the detected maser emission in IRAS06099$+$1800 and G10.62$-$0.38 was turned out to be toward the radio continuum peaks. These two are thus included in the list of detections, while the other four are excluded (see marks and notes in Table\,\ref{tb:source}).


\section{RESULTS}\label{sec:result}

We detected 63 (61) \water\ and 41 (37) \methanol\ maser sources toward 102 \uchiis\ in 2010 (2011), and 29 \water\ and 26 \methanol\ maser sources toward 44 \uchiis\ in 2017 (see Tables\,\ref{tb:source} \& \ref{tb:detection} for details). In the 2017 observations with adjusted coordinates, four \water\ and seven \methanol\ maser sources were newly detected in comparison to the previous epochs. The \water\ and \methanol\ maser emission were at least once detected in 74 and 55 sources, respectively. Forty six sources are associated with the emission of both maser species. Twenty eight sources are associated with only \water\ maser emission, while nine are associated with only \methanol\ maser emission. The present survey detected 3 and 27 new maser sources at 22 and 44,\ghz, respectively. Table\,\ref{tb:detection} summarizes the detection rate of each maser species for different epochs and sub-samples. The overall detection rates of \water\ and \methanol\ maser emission are 72$\pm$9\% and 53$\pm$10\%\footnote{The percentage error is estimated by using the normal approximation formula of the binomial confidence interval with 95\% confidence: $p\pm z_{1-\alpha/2}\sqrt{\frac{p(1-p)}{n}}$, where $p$ and $n$ are the portion of interest and the sample size. $\alpha$ is the coveted confidence, and $z_{1-\alpha/2}$ is the z-score for the coveted confidence level which is 1.96 for 95\% confidence in this paper.}, respectively. These final detection rates exclude the false detections due to offsets between the two coordinates in 2010/2011 and 2017. The false detections are all from three 44\,\ghz\ \methanol\ maser sources: IRAS06058$+$2138, G19.07$-$0.27 and IRAS20264$+$4042 (see the 8th column of Table\,\ref{tb:source}). The WC89a subsample has a bit higher detection rates of both masers in each epoch than those of the KCW94 subsample (see Table\,\ref{tb:detection}). This might be because the former has a brighter range of bolometric luminosities (4.5\,$\leq$\,Log($L_{\rm bol}$)\,$\leq$\,7.0) than the latter  (3.0\,$\leq$\,Log($L_{\rm bol}$)\,$\leq$\,6.5) (Table\,\ref{tb:properties}). We will discuss the relationship between each maser luminosity and the bolometric luminosity of the central object  in \S\,\ref{sec:lbol}.

\subsection{General Properties of Detected \water\ and 44\,\ghz\ \methanol\ Masers}\label{sec:general}

\begin{figure*}
	\centering
	\scalebox{0.19}[0.19]{\includegraphics{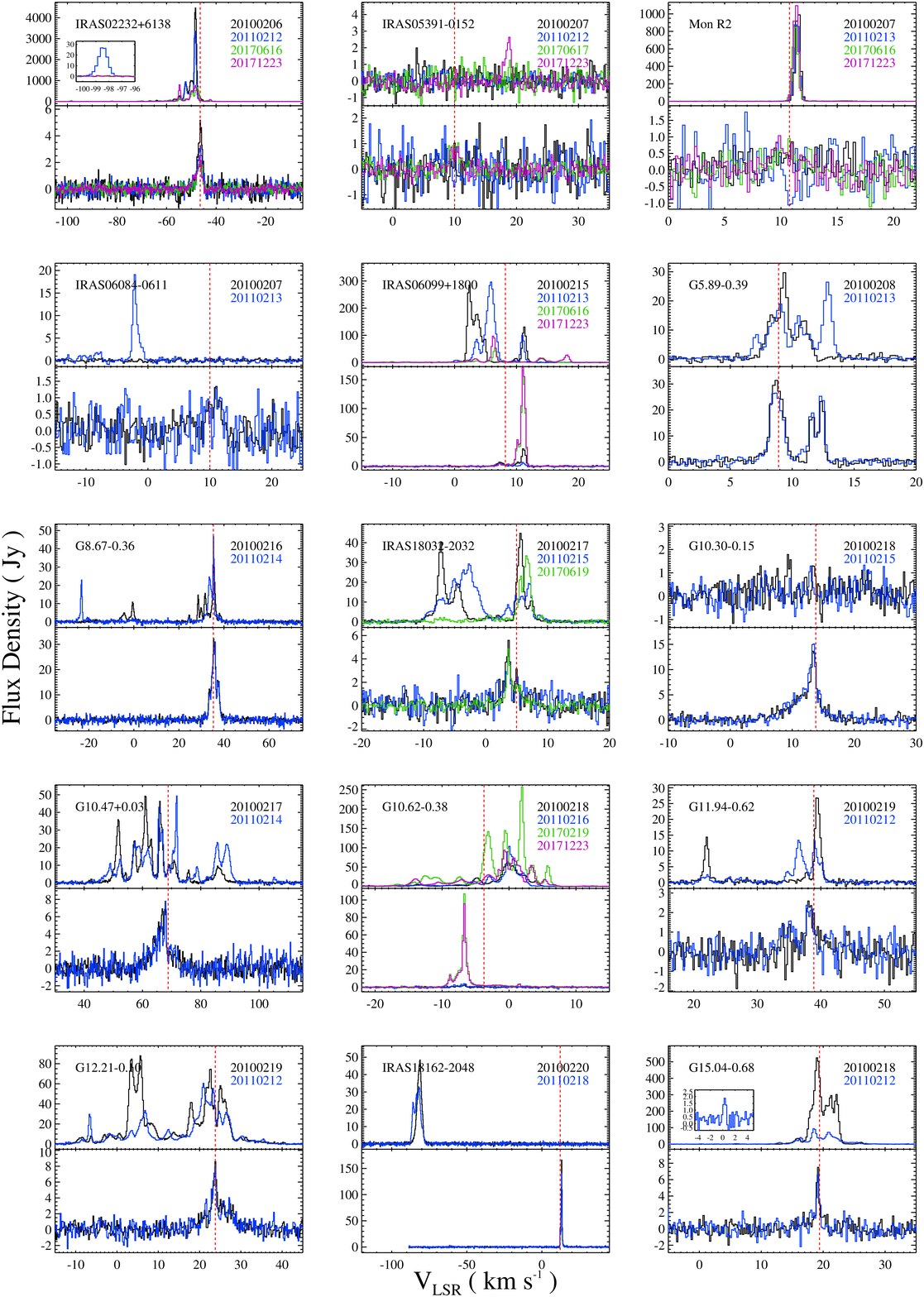}}
	\caption{\label{fig:spec_both}22~\ghz\ \water\ (upper panel) and 44~\ghz\ \methanol\ (lower panel) maser spectra of the sources detected in both maser emission. Different colors indicate different observation epochs. In each panel, the source name is presented at the top-left corner and the observing dates are listed at the top-right corner. The vertical dotted line shows the systemic velocity. High-velocity \water\ maser lines are expanded in the insets. 
	}
\end{figure*}

\begin{figure*}
	\centering
	\scalebox{0.19}[0.19]{\includegraphics{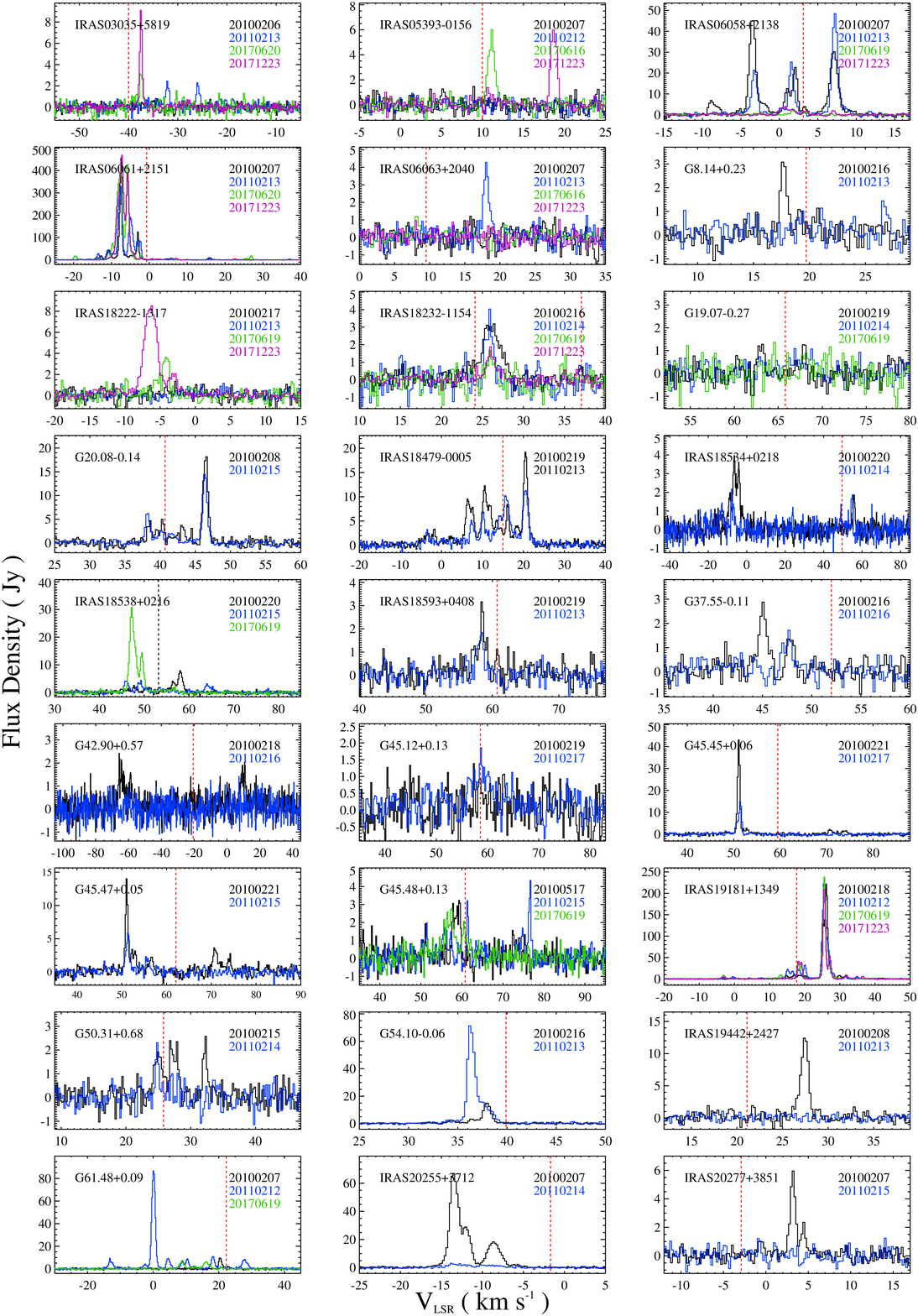}}
    
	\caption{\label{fig:spec_water}Spectra of only the \water\ maser-detected sources. Same symbols as in Figure~1.
	}
\end{figure*}

Figures\,\ref{fig:spec_both} and \ref{fig:spec_water} show the detected \water\ maser spectra for all the epochs. The \water\ maser lines show significant variations in the line profile and flux density between the individual epochs and span a wide range of velocities. Many of the sources show maser peaks at velocities significantly shifted from the systemic velocities marked by the vertical dotted lines. Here, the systemic velocities were obtained from the previous molecular line observations of the parental dense cores \citep{churchwell1990,wouterloot1993,anglada1996,bronfman1996,shepherd1996}. In particular, G30.54$+$0.02, G34.26$+$0.15, IRAS19095$+$0930, and W51d show maser lines shifted by more than 100\,\kms. The sources with high-velocity maser lines will be discussed in more detail in \S\,\ref{sec:high-velo_water}. 

\begin{figure*}
\centering
\scalebox{0.19}[0.19]{\includegraphics{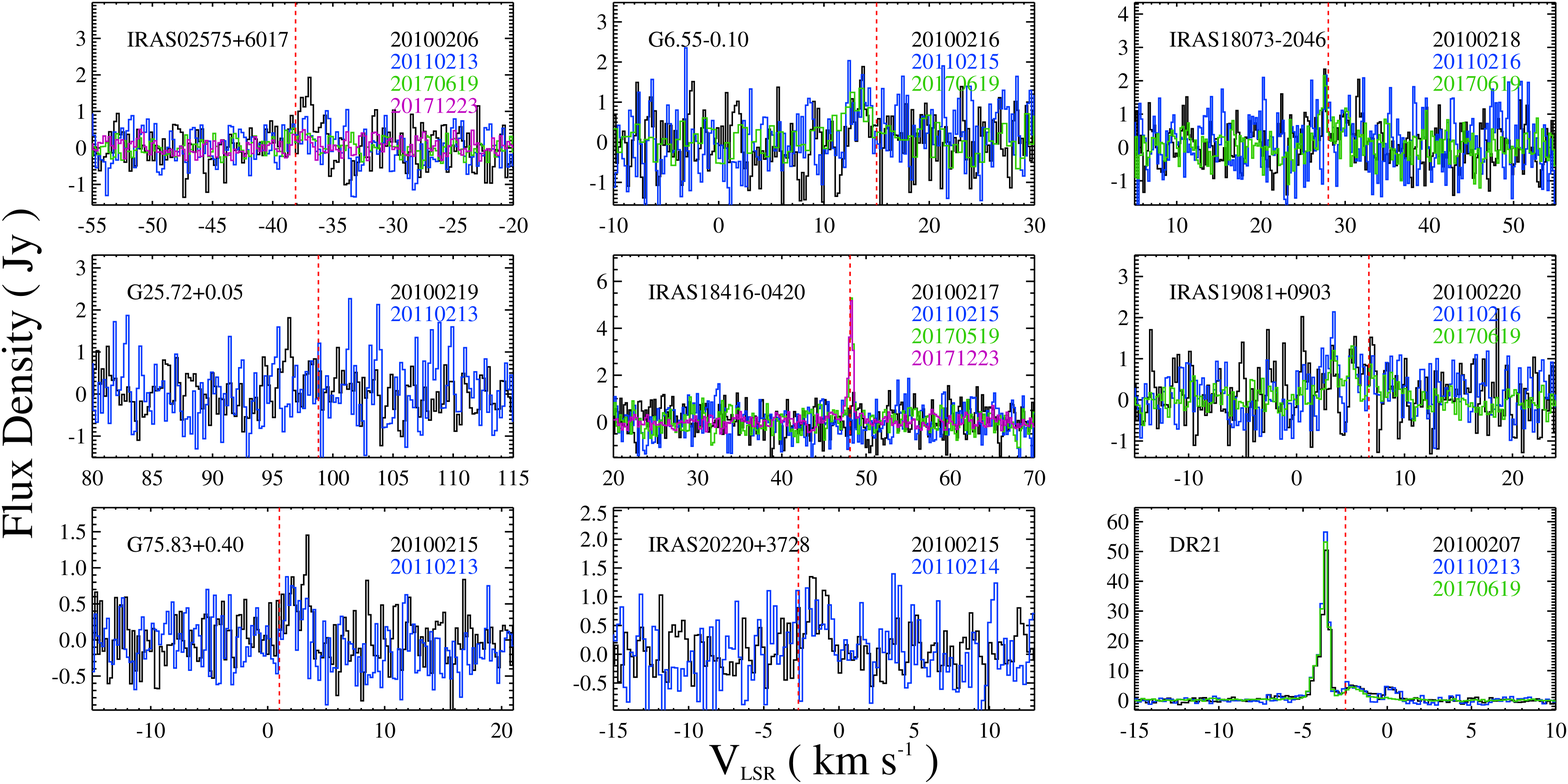}}
\caption{\label{fig:spec_methanol}Spectra of only the \methanol\ maser-detected sources. Same symbols as in Figure~1.}
\end{figure*}

Figures\,\ref{fig:spec_both} and \ref{fig:spec_methanol} exhibit the detected \methanol\ maser spectra. \methanol\ maser sources usually have one or two spectral features. These masers tend to show much simpler line profiles and narrower line-widths than \water\ masers. They always have very similar velocities to those of the natal dense cores. We also detected quasi-thermal \methanol\ emission features along with the maser emission toward some sources: G10.30$-$0.15, G10.47$+$0.03, G12.21$-$0.10, G31.41$+$0.31, IRAS18456$-$0129, W51d, and IRAS19410$+$2336. These sources will be discussed in \S\,\ref{sec:quasi-thermal_methanol}. The line parameters of the detected \water\ and \methanol\ masers are tabulated in Tables\,\ref{tb:water_linepara} and \ref{tb:methanol_linepara}, respectively.

\begin{figure}
\centering
\includegraphics[width=0.49\textwidth]{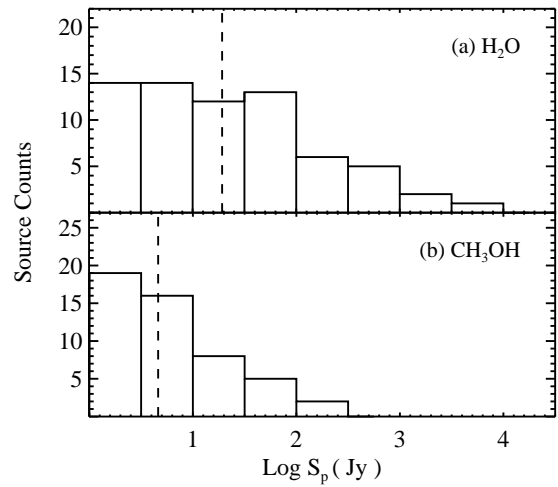}
\caption{\label{fig:hist_flux}Histogram of the flux densities for (a) \water\ and (b) \methanol\ masers. The bin size is 0.5\,dex. The vertical dashed line indicates median value of the flux densities of each maser species.}
\end{figure}

Figure\,\ref{fig:hist_flux} compares the peak flux densities of the detected \water\ and \methanol\ masers. In this comparison, we used the first-epoch data for the sources with the coordinate offsets $<\,10''$ and the third- or fourth-epoch data for the others with larger offsets. The \water\ masers have higher flux densities than the \methanol\ masers in the vast majority of the sources with both masers. The median values of \water\ and \methanol\ masers are $\sim$\,19\,Jy and $\sim$\, 5\,Jy, respectively, which are marked in the black dashed lines. The majority of \water\ maser sources have higher peak flux densities than 10\,Jy, while most \methanol\ maser sources have lower values than that. In addition, the distribution range is much wider for \water\ masers than for \methanol\ masers. Some of the sources, however, show higher peak fluxes of \methanol\ masers than those of \water\ masers: G5.89$-$0.39, G10.30$-$0.15, IRAS18162$-$2048, G23.71$+$0.17, G27.28$+$0.15, G30.54$+$0.02, IRAS18469$-$0132, G33.92$+$0.11, G35.05$-$0.52, G42.42$-$0.27, and IRAS22176$+$2048. Three of them (18162$-$2048, G23.71$+$0.17, G35.05$-$0.52, and IRAS22176$+$6303) have 2$-$16 times stronger \methanol\ maser emission than \water\ maser emission.

\begin{figure}
\centering
\scalebox{0.59}[0.59]{\includegraphics{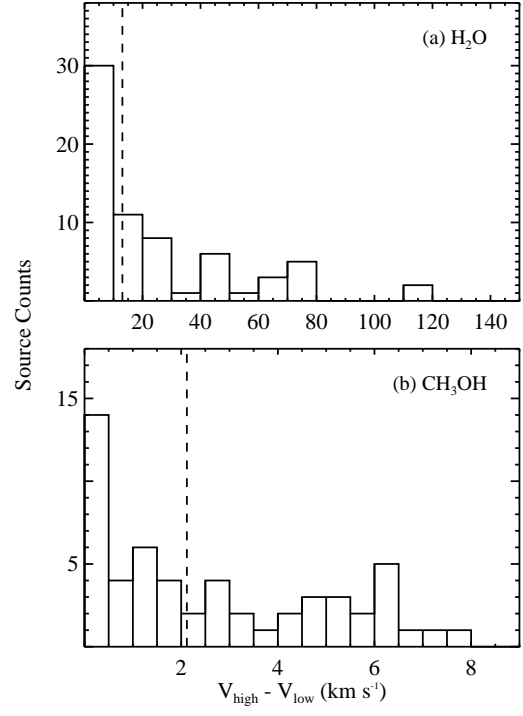}}
\caption{\label{fig:hist_velo}Histogram of the velocity ranges of (a) \water\ and (b) \methanol\ masers. The bin sizes are 10\,\kms\ and 0.5\,\kms\ for \water\ and \methanol\ masers, respectively. The vertical dashed line indicates median value of the velocity ranges for each maser species.}
\end{figure}

Figure\,\ref{fig:hist_velo} displays the velocity range distributions of the detected \water\ and \methanol\ masers. Here $V_{\rm low}$ and $V_{\rm high}$ are the lowest and highest velocities of the velocity range of each source (Table~3). These velocities were measured at the 3$\sigma$ noise level. All the \methanol\ maser sources show smaller velocity ranges than 10\,\kms, and the vast majority (74\%) of them have velocity ranges $<$\,5\,\kms. On the other hand, about (60\%) of the \water\ maser sources have larger velocity ranges than 10\,\kms. Moreover, five sources have velocity ranges greater than 100\,\kms, and many of their maser lines are extremely shifted and variable as seen in the upper panel of Figure\,\ref{fig:spec_both} and Table\,\ref{tb:water_linepara}: G30.54$+$0.02 (111.3\,\kms\ in 2011), G34.26$+$0.15 (114.3\,\kms\ in 2010), IRAS19095$+$0930  (117.8\,\kms\ in 2010), W51d  (223.4\,$-$\,300.6 \,\kms\ in 2010, 2011, and 2017), IRAS20081$+$3122 (115.9 and 104.1\,\kms\ in 2010 and 2011). W51d shows the largest velocity range in each of the three observed epochs: 223.4\,\kms\ in 2010, 250.3\,\kms\ in 2011, 300.6\,\kms\ in 2017 June. The W51d is located together with another \uchii\ (W51d1) and a hot core (W51-North) within a few arcseconds in the W51 IRS2 complex \citep{zapata2009,goddi2015}. Due to our bigger beam size than the complex region, there is an uncertainty to clarify the origin of the extremely-shifted \water\ maser features.

\subsection{Variability of Detected \water\ and 44~\ghz\ \methanol\ Masers}\label{sec:variation}

As seen in Figures\,\ref{fig:spec_both} and \ref{fig:spec_water}, \water\ masers generally show significant variations in the line intensity, profile, and velocity over the one-year time interval of our observations. It is well known that \water\ masers considerably vary on timescales as short as a few weeks (e.g., \citealt{wilking1994, claussen1996, furuya2001, furuya2003, bae2011}). For instance, no \water\ maser was detected toward IRAS19410$+$2336 in 2010, but bright four maser lines, with peak fluxes of 16.2\,$-$\,48.7\,Jy, appeared at velocities between 15.4 and 28.1\,\kms\ in 2011. IRAS20255$+$3712 showed an opposite case that bright three \water\ maser lines immensely weakened (the peak line intensity dropped from 67.1 to 3.1\,Jy) or disappeared from 2010 to 2011. We also found such variations over about six-month time interval toward some of the observed sources in 2017 June and December. For example, one of red- or blueshifted lines of IRAS18456$-$0129 and G43.18$-$0.52 changed by $\sim$4 times in the peak flux. In the case of IRAS05393$-$0156, the detected maser lines have different velocities between the third and fourth epochs. \cite{goddi2007} suggested that such high variability of \water\ maser emission can be caused by outflows or shocks passing through the emitting region. 

In contrast, 44\,\ghz\ \methanol\ masers rarely show such variability (see Figures\,\ref{fig:spec_both} and \ref{fig:spec_methanol}). For instance, G5.89$-$0.39 exhibited the same spectra in the first and second epochs. Furthermore, when the spectra are compared with the one obtained by \cite{bachiller1990}, no significant difference is found between them. However, there are some sources where noticeable variations in the line intensity are observed, i.e., IRAS06058$+$2138, G19.61$-$0.23, IRAS19598$+$3324. For example, G19.61$-$0.23 was found to consist of two \methanol\ maser lines at 41.2\,\kms\ (19.0\,Jy) and 46.1\,\kms\ (3.8\,Jy) in 2010, but only a single line was detected with about half the peak flux density, 8.2\,Jy at 40.9\,\kms\ in 2011. The two \methanol\ lines detected in 2010 were observed in 2017 June, with a very similar profile despite its positional offset of 11$''$. \cite{bachiller1990} detected one maser line with a peak flux density of $\sim$24\,Jy at 41.4\,\kms\ toward this source at the same position. For comparison, 6.7~\ghz\ Class~II \methanol\ masers appear to be considerably more variable  (e.g., \citealt{goedhart2004, durjasz2019}). They show even periodic variations in a few tens of sources (e.g., \citealt{sugiyama2018, olech2019}).

\begin{figure}
\centering
\scalebox{0.56}[0.56]{\includegraphics{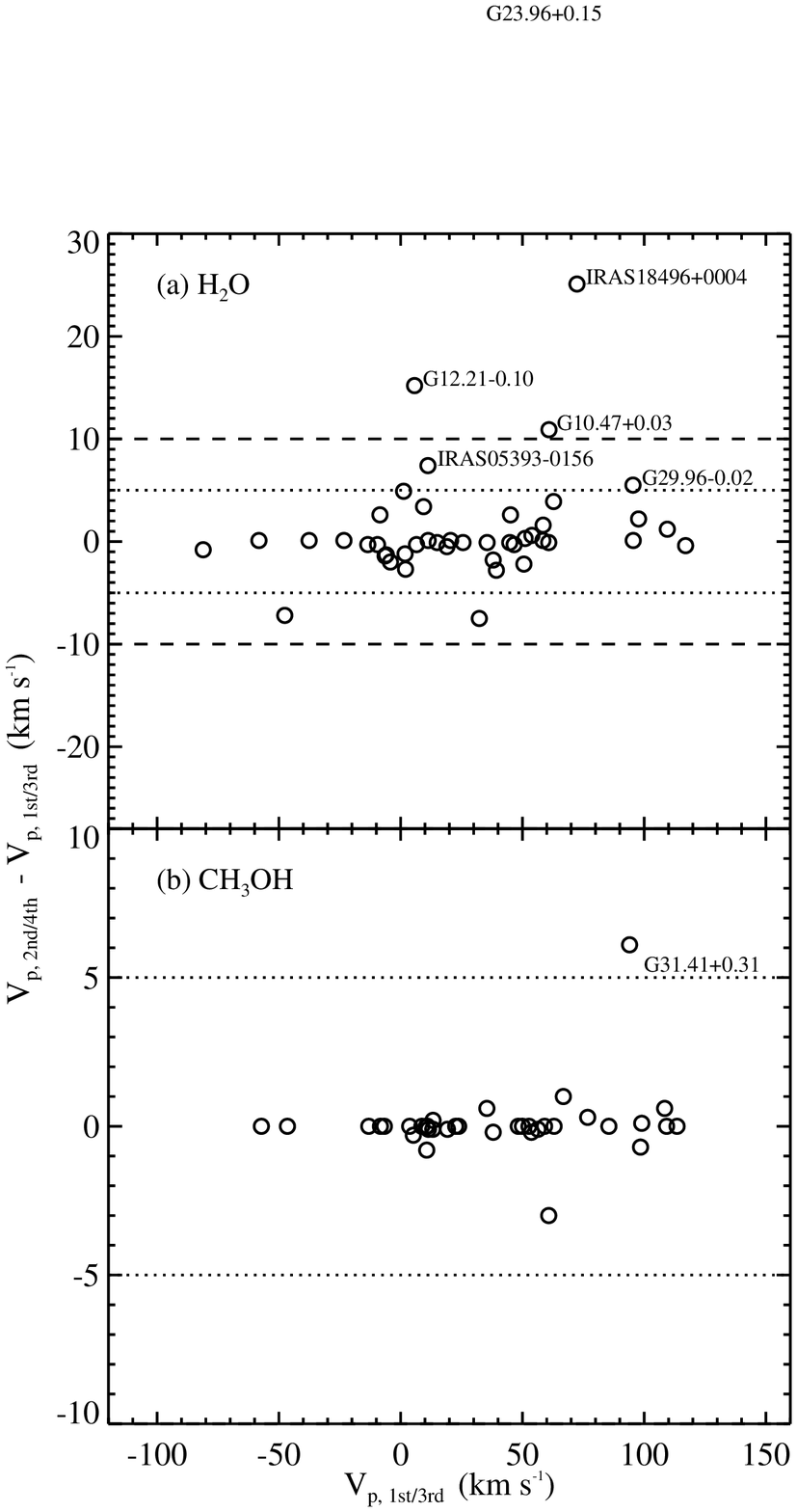}}
\caption{\label{fig:velo_variation}
Peak velocity difference between the first and second or the third and fourth epochs versus the peak velocity of the first or the third epochs for (a) \water\ and (b) \methanol\ masers. G23.96$+$0.15 is not plotted in the upper panel because its velocity difference is in excess of the plot range (see details in the text). }
\end{figure}

Figure \ref{fig:velo_variation} plots the peak velocity differences between the first and second epochs against the peak velocity of the first epoch for the sources with positional offsets $< 10''$, and between the third and fourth epochs for the sources with offsets $> 20''$. The velocity differences are invariably less than 10\,\kms\ for the \methanol\ masers. The standard deviation ($\sigma$) is 1.2\,\kms\ for all the sources, and it reduces to 0.6\,\kms\ without taking into account a large variable source, G31.41$+$0.31, with a velocity difference of 6.1\,\kms. It is worth noting that in G31.41$+$0.31 the difference between the two peak fluxes of the first and second epochs (0.4~Jy) is smaller than the rms noise level, $\sim$0.6~Jy (Table~4). The small standard deviation indicates little variability of the \methanol\ peak velocities over six-month or one-year time intervals. On the other hand, the \water\ masers show much larger variations in the peak velocity. The standard deviations ($\sigma$) are 9.0\,\kms\ for all the sources and 5.3\,\kms\ after excluding G23.96$+$0.15, which shows a velocity change of  51\,\kms. \cite{genzel1977} and \cite{churchwell1990} detected \water\ maser emission around the systemic velocity of 79.6\,\kms\ toward G23.96$+$0.15. We also detected one maser line at a velocity of 82.4\,\kms\ in 2011 but found only new high-velocity maser line at a velocity of 31.6\,\kms\ in 2010. Outflow activities of the same young stellar objects (YSOs) can cause these high variations in the peak velocity, although we cannot exclude the possibility that the detected maser lines in different epochs are associated with distinct YSOs within our beam at 22\,\ghz. In order to address this issue, high-resolution observations with interferometers are required. Despite much larger variations on the whole, 57\% (31/54) of the \water\ maser sources show smaller variations than 2\,\kms. \cite{breen2010a} also reported that a similar fraction (62\%) of their 207 \water\ maser sources showed small velocity changes within $\pm$\,2\,\kms\ for about nine-month time interval. These results suggest that for each of \water\ and \methanol\ masers the same emitting regions continuedly generate bright spectral features in the majority of the maser sources.

\section{Notes on some individual sources}\label{sec:individual}
\subsection{High-velocity \water\ Maser Sources}\label{sec:high-velo_water}

Twenty-four of the 74 \water\ maser-detected sources have maser lines shifted by more than 30\,\kms\ with respect to the systemic velocities (Table\,\ref{tb:high-velo}). Thirteen of them show extremely high-velocity ($>$50\,\kms) components. Out of the 24 sources, 15 (62\%) have only blue-shifted high-velocity components, five (21\%) have both blue- and red-shifted ones, and four (17\%) have only red-shifted ones. In five sources the high-velocity components are the strongest lines: IRAS18162$-$2048, G23.96$+$0.15, G30.54$+$0.02, IRAS18534$+$0218, G42.90$+$0.57 (Figures\,\ref{fig:spec_both} \& \ref{fig:spec_water}). They are all blue-shifted except for G30.54$+$0.02 in the first epoch. As mentioned earlier, \water\ masers are known to trace the accretion disks and/or inner jets/outflows of YSOs (e.g., \citealt{torrelles1997,goddi2011,burns2015}). Since it is very difficult to expect such high-velocity components in accretion disks, they are very likely to originate from jets/outflows. The measured radial velocity depends on the inclination angle ($\theta$) between the jet/outflow axis and the line of sight as cos\,$\theta$. Thus we cannot exclude the possibility that \water\ maser sources with lower relative velocities can also be related to jets/outflows with large inclination angles. Table\,\ref{tb:high-velo} lists the 24 high-velocity \water\ maser sources. In the table and afterwards, HV and EHV mean high-velocity (30$-$50\,\kms) and extremely high-velocity ($>$\,50\,\kms) \water\ maser lines, respectively.

\subsubsection{G8.67$-$0.36}
G8.67$-$0.36 shows blue-shifted HV and EHV components with respect to the systemic velocity (35.3~\kms) in the first and second epochs, respectively (Figure\,\ref{fig:spec_both}). The EHV component at a velocity of $-$23.2\,\kms\ is detected in this study for the first time, while the HV component at a velocity of $-$4.3\,\kms\ has been reported by previous single-dish and VLA surveys \citep{forster1989, churchwell1990,hofner1996}. There is another massive YSO (G8.68$-$0.37) close to G8.67$-$0.36, but it is located about 1$'$ away and has a systemic velocity of $+$37.2\,\kms\ \citep{longmore2011}. It is unclear whether G8.68$-$0.37 is associated with \water\ maser, although \cite{valdettaro2001} detected one maser line at 33.2\,\kms\ toward a midpoint between the two objects using the Medicina 32 m telescope with a beam size (FWHM) of 1.9$'$. Thus the HV and EHV components are likely associated with jets/outflows from G8.67$-$0.36.

\subsubsection{IRAS18162$-$2048}
In the 2010 and 2011 epochs, IRAS18162$-$2048 (also known as G10.8841$-$2.592, HH~80$-$81, or GGD~27) shows strong maser lines around a velocity of $-$80\,\kms, which are EHV blueshifted components with respect to the systemic velocity (12.2~\kms) (Figure\,\ref{fig:spec_both}). These EHV components between velocities of $-$90 and $-$40\,\kms\ have been reported by several previous surveys \citep{gomez1995,codella1996,marti1999,furuya2003,kurtz2005}. Other maser lines around the systemic velocity have also been reported by \citet{furuya2003} and \citet{kurtz2005}, although none of those components was detected in our observations. According to the VLA observations of \citet{gomez1995} and \citet{kurtz2005}, the EHV blueshifted components are located about 7$''$ northeast of the \uchii\ and thermal radio jet, and thus we cannot exclude the possibility that they may be related to a separate YSO.

\subsubsection{G30.54$+$0.02}
G30.54$+$0.02 shows only redshifted EHV lines at a velocity of $\sim$102\,\kms\ in 2010 (Figure\,\ref{fig:spec_both}). The EHV lines are shifted from the systemic velocity (48.0\,\kms) by $\sim$54\,\kms. The redshifted components disappeared in 2011, and new weak lines appeared between $-$66 and $-$33~\kms\ and near the systemic velocity. \cite{urquhart2011} also detected multiple maser lines between $-$40 and +70~\kms\ using the Green Bank 100~m telescope (FWHM=30$\arcsec$) in a similar period (from 2009 Nov 25 to 2010 Dec 10). Hence the blue- and red-shifted EHV components seem to be very variable. There is no previous report about the redshifted EHV components.

\subsubsection{IRAS20081$+$3122}
IRAS20081$-$3122 (ON 1) shows a bunch of maser lines from $-$70 to $+$60\,\kms, including blueshifted HV/EHV and redshifted HV components, in 2010 and 2011 (Figure\,\ref{fig:spec_both}). The blueshifted components are stronger than the redshifted ones. According to the \water\ maser monitoring result of \cite{felli2007} toward this object for two decades (from 1987 March to 2007 February), maser lines around the systemic velocity (11.6~\kms) were usually the strongest, and blue- and redshifted HV/EHV lines were more frequently variable.

\subsubsection{G75.78$+$0.34}
G75.78$+$0.34 (ON 2) shows multiple maser lines between $-$25 and $+$20\,\kms\ in 2010 and 2011 (Figure\,\ref{fig:spec_both}). We detected a weak redshifted HV component at a velocity of $\sim$44.5~\kms\ only in 2010. The HV line has not been detected either in the \water\ maser monitoring observations of \cite{lekht2006} from 1995 to 2004 or in the VLA observation of \cite{hofner1996} due to their short velocity coverages. In the VLA image of \cite{hofner1996} a cluster of \water\ maser features with velocities between $-$20 and $+$15\,\kms\ are located $\sim2\arcsec$ from the front of this cometary \uchii.

\subsection{44~\ghz\ \methanol\ Quasi-thermal Emission Sources}\label{sec:quasi-thermal_methanol}

As mentioned in \S\,\ref{sec:general}, the detected 44~\ghz\ \methanol\ maser lines usually have line-widths $\lesssim$1\,\kms, much smaller than typical molecular line-widths, and they mostly have narrower widths than \water\ maser lines in high-mass star-forming regions. However, broad \methanol\ line wings are detected toward 7 sources with our sensitivity: G5.89$-$0.39, G10.30$-$0.15, G10.47$+$0.03, G12.21$-$0.10, G31.41$+$0.31, IRAS06058+2138, IRAS18456$-$0129, W51d, IRAS19410+2336. These features seem to be quasi-thermal emission rather than maser emission. 

Several previous surveys of Class\,I \methanol\ masers have reported the detection of quasi-thermal emission accompanying maser emission \citep{bachiller1990,haschick1990,slysh1994,mehringer1997, pratap2008, kalenskii2010, kim2018a}. \cite{haschick1990} detected quasi-thermal emission from a few sources, including Orion-KL and Sgr\,A, among a half of 50 galactic star-forming regions. Also, \cite{pratap2008} detected 36\,\ghz\ (4$_{-1}-$3$_0$ E) \methanol\  quasi-thermal emission from NGC\,7538, and \cite{kalenskii2010} found quasi-thermal \methanol\ emission at 36, 44, and/or 95\,\ghz\ frequencies toward some low-mass YSOs.


\section{DISCUSSION}\label{sec:discussion}

\subsection{Detection Rates}\label{sec:detection}
We detected 22\,\ghz\ \water\ and 44\,\ghz\ \methanol\ maser emission toward 72\% and 53\% of the observed 103 \uchiis, respectively. These detection rates are determined from the combined result of all epochs and are mostly higher than those of each epoch, e.g., 62\% and 40\% in the first epoch, due to time variability of maser emission (Table\,\ref{tb:detection}). For comparison, we investigate the detection of 6.7~\ghz\ Class~II \methanol\ maser emission toward the sources in our sample from the literature \citep{xu2003,pandian2007,caswell2010,green2010,green2012,szymczak2012,hou2014,breen2015,hu2016}. At least 34\% (35) of them seem to be related to the maser emission, assuming a matching radius of 5$''$ (see Table\,\ref{tb:source}). Eighty sources in our sample were covered by the Methanol Multibeam (MMB) survey \citep{green2009} using the Parks 64 m telescope \citep{caswell2010,green2010,green2012,breen2015}. The detection rate is estimated to be 30\% (24/80), considering 24 out of the 35 sources are distributed in the MMB survey area (see Table\,\ref{tb:source}).

To understand the relationship between maser activity and central objects, we investigate the \water\ maser detection rates of massive star-forming regions in different evolutionary stages, including infrared dark cloud cores (IRDCs), HMPOs, and \uchiis. The \water\ maser detection rate of \uchiis\ in this survey is significantly higher than the detection rates of HMPOs (42\%, \citealt{sridharan2002}; 52\%, \citealt{szymczak2005}; 52\%, \citealt{urquhart2011}; 51\%, \citealt{kang2015}; 45\%, \citealt{kim2018a}) and IRDCs (12\%, \citealt{wang2006}; 35\%, \citealt{chambers2009}) although these surveys were done with similar or better sensitivities of $\sim$0.1$-$0.5~Jy. This suggests that the detection rate of \water\ maser increases with the evolution of the central objects. This trend is in contrast with the survey results toward low- and intermediate-mass star-forming regions in which the detection rate rapidly decreases as the central objects evolve. \cite{furuya2001} detected \water\ maser emission in 40\% of Class~0, 4\% of Class~I, and none of Class~II objects in low-mass star-forming regions. \cite{bae2011} also found a similar trend toward intermediate-mass YSOs: 50\% for Class~0, 21\% for Class~I objects, 3\% of Herbig Ae/Be stars. This difference might be caused by different environments surrounding low- and high-mass YSOs, as \cite{bae2011} suggested. The circumstellar materials are mostly dispersed by protostellar outflows around low- and intermediate YSOs in later evolutionary stages. On the other hand, \uchiis\ are still deeply embedded in the parental molecular clouds due to their faster evolution and much larger amount of ambient matter although the ionizing stars have already reached the main-sequence stage.

The situation is very similar for the 44\,\ghz\ \methanol\ maser detection rate. Our value for \uchiis\ is considerably higher than the rate (31$\pm$5\%) toward HMPOs reported by the single-dish survey of \citet{fontani2010} with a comparable sensitivity of $\sim$0.6\,Jy at velocity resolution of $\sim$0.2\,\kms\ to this survey. Their sample consists of two groups: {\em low} and {\em high} sources. The {\em high} sources satisfy the $IRAS$ color criteria of \cite{wood1989b} for \uchiis, i.e., \uchii\ candidates without detectable radio continuum emission in single-dish observations, and are believed to be more evolved than the {\em low} sources. They estimated the detection rates for the two groups separately and found that the rate for the {\em high} sources (48$\pm$8\%) is almost three times higher than that of the {\em low} sources (17$\pm$5\%). The former rate is in a good agreement with ours. \cite{kang2015} and \cite{kim2018a} also obtained lower detection rates of 32\% and 28\%, respectively, with almost the same sensitivities at the same velocity resolution as this survey. On the contrary, the detection rate of 44~\ghz\ \methanol\ maser emission decreases with the evolution of the central objects in intermediate-mass star-forming regions: 36\% for Class\,0, 21\% for Class\,I, 1\% for HAeBe objects \citep{bae2011}. This difference can also be understood by the same explanations as for the \water\ maser detection rate. 

Taking into account low angular resolutions of this survey, we cannot exclude the possibility that the detected masers may be associated with nearby YSOs rather than target \uchiis. However, previous high-resolution VLA observations showed strong angular correlations between the two maser species and \uchiis\ \citep{hofner1996,kurtz2004,gomez2010,gomez-ruiz2016}. These studies found that \water\ maser spots are located within $\sim$0.1\,pc of the \uchiis, and 44\,\ghz\,\methanol\ maser spots are separated by $<$0.5\,pc with a mean separation of 0.2\,pc. \cite{gomez-ruiz2016} found more than twice higher detection rate (63\%) for the $high$ sources than that (27\%) for $low$ sources in their VLA observations of 44\,\ghz\,\methanol\ masers toward the sample of \cite{fontani2010}. As mentioned in \S~1, these two maser species are known to be closely related to protostellar outflows although they trace different parts of the outflows. According to \cite{shepherd1996}, the vast majority of \uchiis\ (90\%, of 94 sources) still have bipolar outflows as in HMPOs \citep{sridharan2002,zhang2005,kim2006}. Therefore, it is not surprising that a significant fraction of \uchiis\ are associated with \water\ and/or 44\,\ghz\,\methanol\ maser emission. Moreover, a few high-resolution observations with the Austrailia Telescope Compact Array (ATCA) provided a hint that 44\,\ghz\,\methanol\ masers can be generated in the shocked regions by the expansion of \uchiis\ (\citealt{voronkov2010,voronkov2014}; see also \citealt{kurtz2004, gomez-ruiz2016}).

\subsection{Relative Velocities of Masers}\label{sec:velocity}

\begin{figure*}
\centering
\scalebox{0.35}[0.35]{\includegraphics{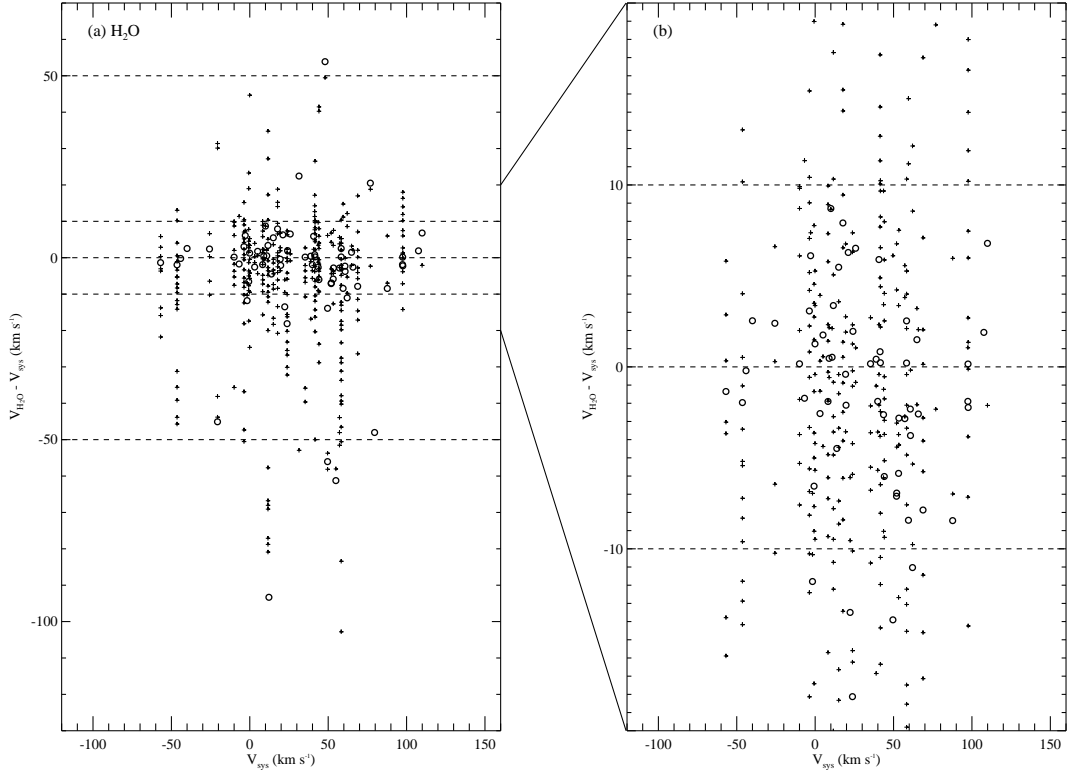}}
\caption{\label{fig:vel_water_mol}
Relative velocity of \water\ maser line versus systemic velocity. The first-epoch data are plotted for the sources with the coordinate offsets $<$10$''$ while the third- or fourth-epoch data are plotted for the other sources. Open circles indicate the strongest maser lines in the individual sources. All the line components of W51d are not plotted here because of its large velocity range, $-$130 to 170\,\kms.} 
\end{figure*}

\begin{figure}
\centering
\scalebox{0.39}[0.39]{\includegraphics{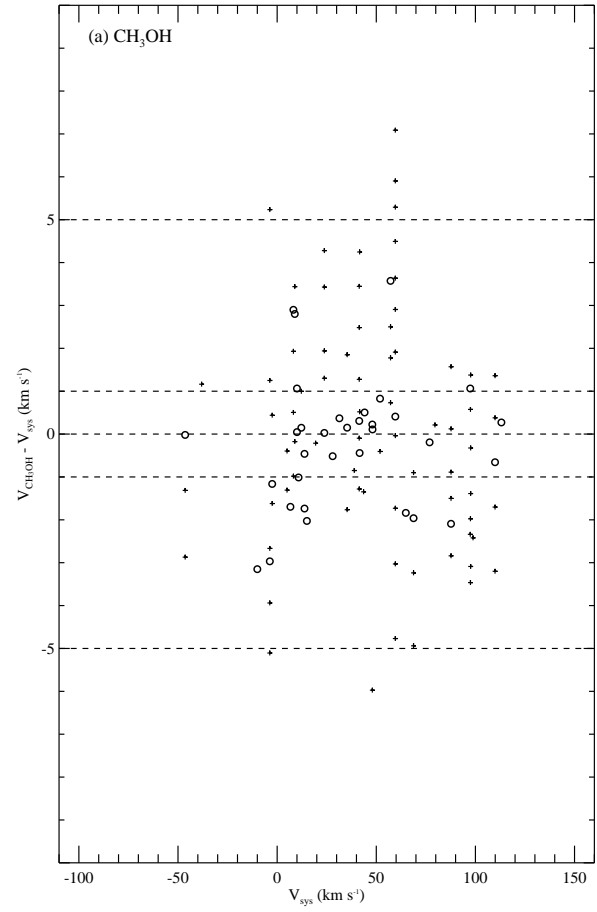}}
\caption{\label{fig:velo_methanol_mol}Relative velocity of \methanol\ maser line versus systemic velocity. Same symbols as in Figure\,\ref{fig:vel_water_mol}. }
\end{figure}

Figures\,\ref{fig:vel_water_mol} and \ref{fig:velo_methanol_mol} show the relative velocities of \water\ and 44\,\ghz\,\methanol\ masers versus the systemic velocities, respectively. In all analyses in this section, we use the 2010 data for the sources with coordinate offsets $<$10$''$ and the 2017 data for the other sources. The relative velocities of \water\ maser lines span from $-$130 to 170\,\kms, including W51d that is not plotted in Figures\,\ref{fig:vel_water_mol} for clarity. The mean value and standard deviation ($\sigma$) of the relative velocities of all components are $-$5.2 and 32.4\,\kms, respectively. The vast majority (88\%) of the peak maser velocities, marked by open circles, are concentrated within the relative velocity range of $\pm$\,20\,\kms. The mean and standard deviation ($\sigma$) of the peak relative velocities are $-$4.4 and 19.2\,\kms, respectively. This relatively good agreement between the peak \water\ maser and systemic velocities corresponds with the results of some previous surveys (e.g., $-$4.5\,\kms, \citealt{kurtz2005}; $-$3.8\,\kms, \citealt{urquhart2011}). Low- and high-velocity \water\ maser components are considered to be produced by slow wide-angle outflows and fast well-collimated jets, respectively \citep{gwinn1992,torrelles2011,goddi2011}. According to theoretical models, \water\ masers are formed in warm ($\sim$\,500\,K) and very dense ($\sim$\,10$^{9}$\,cm$^{-3}$) shocked gas behind fast-velocity ($>$\,50\,\kms) dissociative shocks (J$-$type) or slow-velocity ($<$\,50\,\kms) non-dissociative shocks (C$-$type) \citep{elitzur1989,kaufman1996}. \water\ masers can, therefore, have a wide range of relative velocities because of being accelerated by various velocity outflows.

In contrast to \water\ masers, the relative velocities of 44\,\ghz\,\methanol\ maser lines never exceed $\pm$10\,\kms\ as seen in Figure\,\ref{fig:velo_methanol_mol}. This result supports that \methanol\ molecules can survive only in slow shocks below 10\,\kms\ but are easily destroyed by fast-moving shocks over 10\,\kms\ (e.g., \citealt{garay2002}). The peak velocity components of 44\,\ghz\,\methanol\ masers are also located in a narrow range of relative velocities. They are mostly (79\%) located within $\pm$\,2\,\kms\ with respect to the systemic velocity. This excellent agreement between the peak 44\,\ghz\,\methanol\ maser and systemic velocities is consistent with the fact that Class\,I \methanol\ masers are located in the interacting regions between outflows and the ambient gas \citep{plambeck1990,kurtz2004}. Theoretical models and high-resolution observations have revealed that Class\,I \methanol\ masers originate from cooler and less dense gas (temperature of $\sim$100\,K and densities $\sim$\,10$^{5}$\,cm$^{-3}$) than \water\ masers \citep{cragg1992}. Such condition is found in post-shock regions farther from a central object (0.2$-$0.5\,pc; \citealt{kurtz2004,gomez2010}) than that typically found for \water\ masers ($\sim$\,0.1\,pc; \citealt{hofner1996}). However, it should be noted that in a few sources Class\,I \methanol\ masers appear to be associated with the expansion of \hii\ regions \citep{voronkov2010,araya2008} or by accretion shocks \citep{kurtz2004}. Although \water\ and 44\,\ghz\,\methanol\ masers are commonly excited by collisional pumping by shocks from jets/outflows, there is a significant difference between the velocity distributions of the two maser species. This may indicate that \water\ and \methanol\ masers are generated from different parts of the outflows and are sensitive to different physical conditions in the vicinity of (proto)stars.

\subsection{Environmental Conditions of Masers}\label{sec:environment}

The total isotropic luminosities of \water\ and \methanol\ masers ($L_{\rm{H_{2}O}}$ and $L_{\rm{CH_{3}OH}}$) can be calculated by the following equations:

\begin{eqnarray}
L_{\rm{H_2O}} & = &4 \pi d^{2} {\frac{\nu}{c}} \int{F_{\nu}}\,dv 
\nonumber \\
& = & 2.32 \times 10^{-8} L_{\odot} \bigg(\frac{d}{1~\rm{kpc}}\bigg)^2 \bigg(\frac{\int{S_{\nu}}\,dv}{1\,\rm{Jy\,km\,s}^{-1}}\bigg) \bigg(\frac{\eta_{a}}{0.72}\bigg)^{-1},
\end{eqnarray}

and

\begin{eqnarray}
L_{\rm{CH_3OH}} = 
4.60 \times 10^{-8} L_{\odot} \bigg(\frac{d}{1~\rm{kpc}}\bigg)^2 \bigg(\frac{\int{S_{\nu}}\,dv}{1\,\rm{Jy\,km\,s}^{-1}}\bigg) \bigg(\frac{\eta_{a}}{0.69}\bigg)^{-1}.
\end{eqnarray}

\noindent Here $d$ is the distance to the source, $\nu$ is the observing frequency, $\int{S_{\nu}}\,dv$ is the integrated flux density, and $\eta_{a}$ is the aperture efficiency of the telescope. We adopt the kinematic distances from WC89a and KWC94 and a few other references, as noted in  Table\,\ref{tb:properties}. In a case where the maser emission consists of multiple lines, the sum of the individual components is used for $\int{S_{\nu}}\,dv$ as in some previous studies (e.g., \citealt{furuya2003,bae2011}). The estimated maser luminosities of \water\ and \methanol\ masers are given in columns (8) and (9) of Table\,\ref{tb:properties}, respectively. The table also provides the physical parameters of \uchiis\ and their parental clumps.

\subsubsection{Comparison of Bolometric Luminosity and Maser Luminosity}\label{sec:lbol}

\begin{figure}
\centering
\scalebox{0.59}[0.59]{\includegraphics{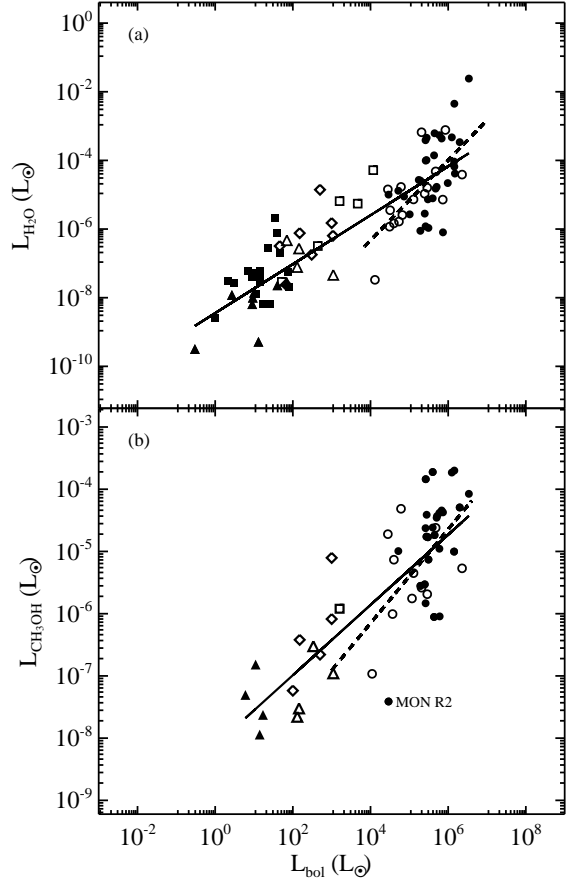}}
\caption{\label{fig:lmaser_lbol}
Isotropic maser luminosity as a function of bolometric luminosity of the central star for (a) {\water} and (b) {\methanol} masers. Filled circles are WC89a sources while open circles are KCW94 sources. Open diamonds, open triangles, and open squares indicate IMYSOs in Class\,0, Class\,I, and HAeBe stages, respectively \citep{bae2011}. In (a), filled triangles and filled squares represent LMYSOs \citep{furuya2003}. In (b), filled triangles indicate LMYSOs \citep{kalenskii2013}.
The dotted and solid lines are the best-fit relations for \uchiis\ and for all objects, respectively.}
\end{figure}

Figure\,\ref{fig:lmaser_lbol} shows $L_{\rm{H_{2}O}}$ and $L_{\rm{CH_{3}OH}}$ against the bolometric luminosity of the central star ($L_{\rm bol}$). The relation for \water\ masers by linear least-squares fitting is  $L_{\rm{H_{2}O}}=1.01\times10^{-11}(L_{\rm{bol}})^{1.17}$ with a correlation coefficient ($\rho$) of 0.66. The fitted result for 44\,\ghz\,\methanol\ masers is $L_{\rm{CH_{3}OH}}=6.70\times10^{-10}(L_{\rm{bol}})^{0.76}$ with $\rho=$0.52. The former and latter relations are shown by dotted lines in the upper and lower panels, respectively. Both $L_{\rm{H_{2}O}}$ and $L_{\rm{CH_{3}OH}}$ tend to increase with $L_{\rm bol}$. The correlations get significantly stronger when the data points of the low- and intermediate-mass regimes are added to the \uchiis. The newly fitted relations are $L_{\rm{H_{2}O}}=3.56\times10^{-9}(L_{\rm{bol}})^{0.71}$ with $\rho=0.89$, and $L_{\rm{CH_{3}OH}}=7.61\times10^{-9}(L_{\rm{bol}})^{0.56}$ with $\rho=0.80$. They are shown by solid lines in both panels. Here the data of LMYSOs are taken from \cite{furuya2003} for \water\ masers and \cite{kalenskii2010} for \methanol\ masers, and the data of IMYSOs are from \cite{bae2011} for both maser species. \cite{bae2011} also did similar analyses by combining their data of IMYSOs with the preliminary results of this study and obtained a little bit higher slopes for both relations, 0.84 and 0.73. This is mainly because they derived $L_{\rm bol}$ of \uchiis\  using the VLA radio continuum data rather than the $IRAS$ data as in this study. For a given \uchii, the $L_{\rm bol}$ from the radio continuum data is always smaller than that from the $IRAS$ data due to the absorption of part of UV photons by dust inside the \uchii\ \citep{wood1989a,kurtz1994}, and the contributions of low/intermediate-mass star clusters in the region \citep{kim2018b}. These results demonstrate that more luminous (and massive) stars are likely to generate more luminous \water\ and \methanol\ masers.

\begin{figure}
\centering
\scalebox{0.59}[0.59]{\includegraphics{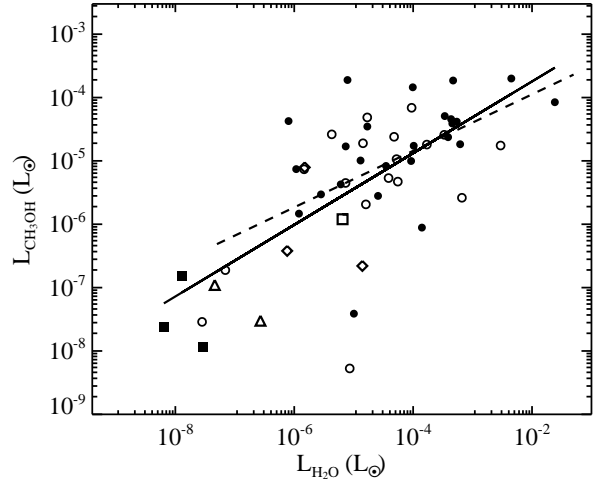}}
\caption{\label{fig:lum_water_lum_methanol} \water\ maser luminosity as a function of \methanol\ maser luminosity. The filled circles are WC89a sources while the open circles are KCW94 source. Diamonds, triangles, squares indicate the data points of IMYSOs in Class 0, Class I, and HAeBe stages from \cite{bae2011}, respectively. 
Filled squares are LMYSOs from \cite{kalenskii2010}. The dotted line shows the fitted relation of only \uchii\ data points, whereas the solid line represents the one of all data points (see the text for more details).}
\end{figure}

Figure\,\ref{fig:lum_water_lum_methanol} shows the comparison of the two maser luminosities. The dotted line is the best fit of only \uchiis\ while the solid line is the fit to all the data points including LMYSOs and IMYSOs. The former is $L_{\rm{CH_{3}OH}}=8.75\times10^{-3}(L_{\rm{H_{2}O}})^{0.45}$ with $\rho=0.54$, and the latter is $L_{\rm{CH_{3}OH}}=2.42\times10^{-3}(L_{\rm{H_{2}O}})^{0.57}$ with $\rho=0.70$. For all the data points, a good correlation is found between the two maser luminosities. This strongly supports the previous suggestion that the two masers are both collisionally pumped by outflows (e.g., \citealt{kurtz2004}). Indeed, \cite{felli1992} found in 56 CO outflow sources that $L_{\rm{H_{2}O}}$ correlates with the mechanical luminosity of the outflow, which is, in turn, proportional to the far-infrared luminosity. \cite{gan2013} detected 95~GHz class~I \methanol\ maser emission toward 62 CO outflow sources and found significant correlations between the maser luminosity and the outflow properties such as mass, momentum, and mechanical energy. \cite{kim2018a} showed from a simultaneous surveys of 44 and 95~GHz \methanol\ masers that the two maser transitions have tight correlations in the line velocity, flux density, and profile (see also \citealt{kang2015}).

\subsubsection{Comparison of Radio Continuum Luminosity and Maser Luminosity}\label{sec:radio_conti}

\begin{figure}
\centering
\scalebox{0.59}[0.59]{\includegraphics{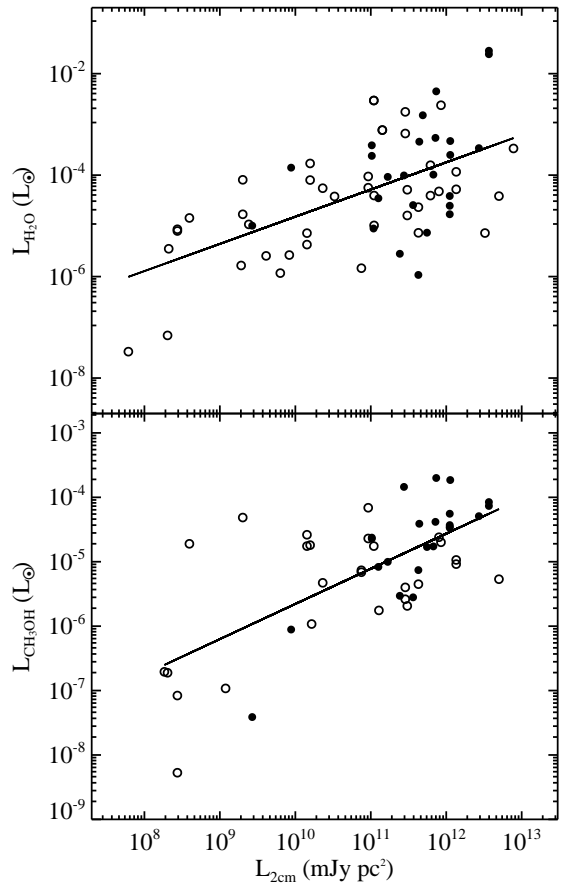}}
\caption{\label{fig:2cm_lmaser}Isotropic maser luminosity as a function of 2\,cm radio continuum luminosity for (a) {\water} and (b) {\methanol} masers.
Filled circles are WC89a sources, and open circles are KCW94 sources.}
\end{figure}

\begin{figure}
\centering
\scalebox{0.59}[0.59]{\includegraphics{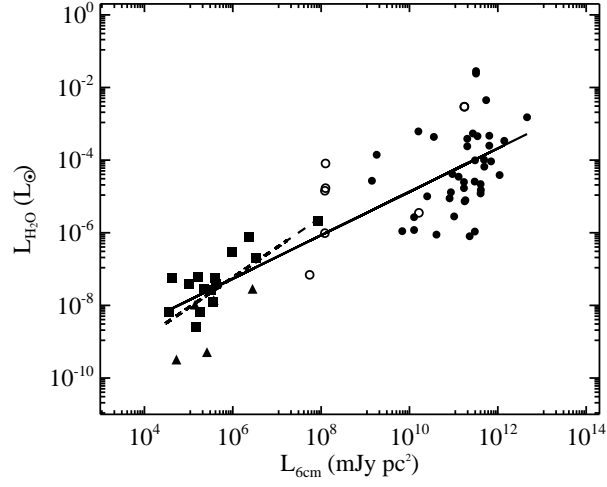}}
\caption{\label{fig:6cm_lum_water}Isotropic \water\ maser luminosity as a function of 6\,cm radio continuum luminosity. Circles represent \uchiis, and filled triangles and squares show LMYSOs from \cite{furuya2003}}
\end{figure}

Figure\,\ref{fig:2cm_lmaser} plots $L_{\rm{H_{2}O}}$ and $L_{\rm{CH_{3}OH}}$ versus the radio continuum luminosity at 2\,${\cm}$ ($L_{\rm{2\,\cm}}$).
The linear least-squares fittings yield $L_{\rm{H_{2}O}}=6.65\times10^{-11}(L_{\rm 2\,{\cm}})^{0.54}$ with $\rho=0.57$ and  $L_{\rm{CH_{3}OH}}=7.71\times10^{-12}(L_{2\,{\cm}})^{0.55}$ with $\rho=0.68$. The $L_{\rm{2\,\cm}}$ also tends to increase with $L_{\rm{H_{2}O}}$ and $L_{\rm{CH_{3}OH}}$ like $L_{\rm{bol}}$. The $L_{\rm{CH_{3}OH}}$ shows a better correlation with $L_{\rm{2\,\cm}}$ than $L_{\rm{H_{2}O}}$. Figure\,\ref{fig:6cm_lum_water} shows a comparison between the $L_{\rm{H_{2}O}}$ and 6\,${\cm}$ radio continuum luminosity ($L_{\rm{6\cm}}$) of LMYSOs \citep{furuya2003} and \uchiis. The dotted line indicates the fitted relation of LMYSOs. On the other hand, the solid line is a leaner-fitting on both LMYSOs and \uchiis, $L_{\rm{H_{2}O}}=1.42\times10^{-11}(L_{\rm 6\,{\cm}})^{0.60}$ with $\rho=0.84$. This seems to be the best correlation between the maser and radio continuum luminosities. Here it should be noted that LMYSOs and \uchiis\ have different origins of radio continuum emission. The radio continuum emission of LMYSOs originates from ionized jets or ionized gas regions by a neutral jet-induced shock \citep{rodriguez1989,meehan1998}, while the radio continuum emission of massive star-forming regions mostly emanates from \hii\ regions. Nonetheless, their $L_{\rm 6 cm}$ have excellent correlations with $L_{\rm{H_{2}O}}$. The reason is likely that the maser luminosity and radio continuum luminosity are both influenced by the bolometric luminosity of the central (proto)star. In fact, the radio continuum luminosity is known to correlate with the bolometric luminosity for LMYSOs \citep{anglada1995,claussen1996,meehan1998}.

\subsubsection{Comparison of Molecular Clump Properties and Maser Luminosity}\label{sec:molecular_core}

\begin{figure}
\centering
\scalebox{0.59}[0.59]{\includegraphics{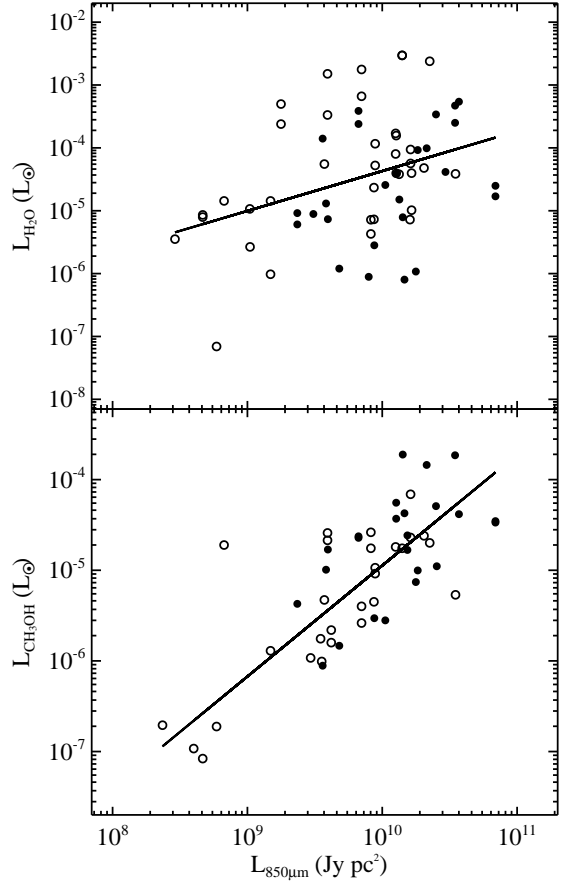}}
\caption{\label{fig:dust_lmaser}Isotropic maser luminosity as a function of 850\,${\mu}$m luminosity for (a) \water\ and (b) \methanol\ masers. Filled circles are WC89a sources, while open circles are KCW94 sources. The solid lines show the best-fits to the data.}
\end{figure}

Figure\,\ref{fig:dust_lmaser} shows comparisons of $L_{\rm{H_{2}O}}$ and $L_{\rm{CH_{3}OH}}$ versus 850\,$\mu$m continuum luminosity ($L_{\rm{850 \mu m}}$). The submillimeter continuum emission emitted from cold dust indirectly provides mass information of molecular clumps and cores \citep{thompson2006,schuller2009}. The fitted results are $L_{\rm{H_{2}O}}=1.95\times10^{-11}(L_{\rm 850 \mu m})^{0.63}$ with $\rho=0.35$ and $L_{\rm{CH_{3}OH}}=6.48\times10^{-18}(L_{\rm 850 \mu m})^{1.22}$ with $\rho=0.77$. While the $L_{\rm{H_{2}O}}$ is weakly correlated with $L_{\rm{850\mu m}}$, the $L_{\rm{CH_{3}OH}}$ strongly correlates with it. \cite{kim2018a} also investigated the correlations of the two maser luminosities with the host clump mass and reported significant correlations for both masers. \cite{de_villiers2014} showed that the outflow mass and mass-loss rate have clear correlations with the clump mass, and \cite{urquhart2014} found that the bolometric luminosity tightly correlates with the clump mass. These correlations suggest that more massive clumps form more massive (and luminous) stars, which emit more energetic outflows that can generate stronger \water\ and class I \methanol\ masers.

\subsection{Evolutionary Sequence for Masers in Massive Star Formation}\label{sec:evolution}
One major question of the maser study is whether a maser species traces any specific evolutionary stage of massive star formation. As massive (proto)stars evolve, their strong radiation and powerful jets/outflows tremendously impact on the physical, chemical, and dynamical properties of the surrounding materials. The changing physical and chemical conditions in the circumstellar region can lead to the production of different maser species in different evolutionary phases. As mentioned earlier, \water, \methanol\ (Class\,I \& Class\,II), and/or OH masers have been found to be associated with a number of massive star-forming regions in different stages. In many cases multiple different maser species have been detected toward the same object (e.g., \citealt{beuther2002}). This indicates that there may be significant overlaps between different maser phases. \cite{ellingsen2006} found that Spitzer point sources with both Class I and Class II\,\methanol\ masers tend to have redder infrared colors than those with only Class II\,\methanol\ masers and argued that Class\,I \methanol\ masers may trace earlier evolutionary stages of massive star formation than Class\,II \methanol\ (and \water) masers. \cite{ellingsen2007} later proposed an evolutionary sequence of different maser species in which \water\ masers primarily appear in the \uchii\ phase while Class\,I \methanol\ masers appear at earlier evolutionary stages and then mostly disappear as \uchiis\ develop (see also \citealt{breen2010b}). However, the high association rates of both \water\ and 44\,\ghz\ Class\,I \methanol\ masers with the \uchii\ phase in this study appear to cast some doubt on the hypothesis. Furthermore, the increasing detection rate as a function of evolution also indicates that Class\,I \methanol\ masers preferentially trace the later stages of massive star formation.

Since massive stars usually form in cluster rather than in isolation, it is possible that the high detection rates we have found may result from outflows driven by other cluster members with lower masses. However, the correlation of maser luminosity with the bolometric and radio continuum luminosities would make this seem less likely (see Figures\,\ref{fig:lmaser_lbol}, \ref{fig:2cm_lmaser}, and \ref{fig:6cm_lum_water}). It is also possible that the high detection rates may be due to the contribution of nearby HMPOs producing outflows within the telescope beam. However, the possibility could be negligible, as we have discussed in \S\,\ref{sec:detection}, according to positional coincidence between maser features and the ionized gas regions in previous high-resolution observations \citep{hofner1996,kurtz2004,voronkov2010,voronkov2014} and the ubiquity of jets/outflows in \uchiis\ \citep{shepherd1996,hatchell2001,kurtz2004}. In the recent VLA high-resolution observations of \cite{gomez-ruiz2016}, the detection rate (54\%, 13 of 24) of 44\,\ghz\,\methanol\ maser emission toward \uchii\ candidates was higher than that (38\%, 16 of 42) toward HMPOs. This is consistent with the result of this survey even though the two surveys  were carried out with very different angular resolutions. In addition, a few studies \citep{araya2008,voronkov2010,voronkov2014,gomez-ruiz2016} imply that Class\,I \methanol\ masers can be produced by the expansion of \hii\ regions, as well.

Our finding therefore appears to agree with the results obtained from several previous high-resolution studies but disagrees with the maser evolutionary sequence put forward by \cite{ellingsen2007} and \cite{breen2010b}. One possible reason for this disagreement could be that their analyses were restricted to massive YSOs with Class\,II \methanol\ masers and did not include the later \uchii\ region stage and so were focused on a relatively small range of evolution. In contrast, we conducted a simultaneous \water\ and 44\,\ghz\ Class\,I \methanol\ masers toward a large sample of \uchiis\ and provide more reliable information of the \uchii\ phase in evolutionary sequence in terms of maser detection. All the results of the present and several previous studies strongly suggest that the occurrence periods of the two maser species are more closely overlapped with the \uchii\ phase than earlier evolutionary phases.

\section{SUMMARY}\label{sec:summary}

We performed a multi-epoch simultaneous survey of 22\,\ghz\ \water\ and 44\,\ghz\ Class\,I \methanol\ masers toward 103 \uchiis.  The main results 
are summarized as follows.

1. \water\ maser emission was detected in 74 (72\%) sources and  \methanol\ maser emission in 55 (53\%) sources. These high detection rates suggest that the occurrence periods of both masers are significantly overlapped with the \uchii\ phase. This is not consistent with the previous suggestion that Class\,I \methanol\ masers fade out as \uchiis\ develop. By combining this with the results of some previous maser surveys towards IRDCs and HMPOs, furthermore, we found that the two detection rates may increase with the evolution of the central objects and peak in the \uchii\ phase. Among the detected sources, 3 \water\ and 27 \methanol\ maser sources are new detections. The WC89a sample show slightly higher detection rates for both masers than the KCW94 sample.

2. \methanol\ maser lines always have small ($<$10\,\kms) relative velocities  with respect to the ambient dense molecular gas, while \water\ maser lines usually have much larger relative velocities. We found a few tens of \water\ maser sources with high-velocity components. Twenty four sources have \water\ maser lines with relative velocities $>$30\,\kms, and thirteen of them have \water\ maser lines with relative velocities $\ge$50\,\kms. It is worth noting that the strongest maser lines have relative velocities $<$20\,\kms\ for the vast majority (88\%) of the \water\ maser-detected sources.

3. \water\ masers generally have multiple velocity components with wide velocity ranges. They often show significant variations in the line shape, intensity, and velocity over six-months or one-year time intervals although the strongest maser lines of the individual sources mostly display small variations in velocity. In contrast, \methanol\ masers usually have one or two velocity components, and rarely exhibit significant variations in the line shape, intensity, or velocity for the same periods. This difference, together with the difference between the typical relative velocities of the two maser species, is very likely to arise from their different emitting regions.

4. The isotropic luminosities of both masers tend to increase with the bolometric luminosity of the central star. In the case of that data points of low- and intermediate-mass star-forming regions are added, they well correlate with the bolometric luminosity. The linear fitted relations $L_{\rm{H_{2}O}}=3.56\times10^{-9}(L_{\rm bol})^{0.71}$ and 
$L_{\rm{CH_{3}OH}}=7.61\times10^{-9}(L_{\rm bol})^{0.59}$. There is a quite good correlation between the two maser luminosities.

5. The two maser luminosities tend to increase with the radio continuum luminosity. The linear fitted relations are $L_{\rm{H_{2}O}}=6.65\times10^{-11}(L_{\rm{2\,\cm}})^{0.54}$ and $L_{\rm{CH_{3}OH}}=7.71\times10^{-12}(L_{\rm{2\,\cm}})^{0.55}$. The \methanol\ maser luminosity also shows a tight correlation with the 850\,$\mu$m emission luminosity of the associated molecular clump. These correlations suggest that more massive clumps form more massive (and luminous) stars that can generate stronger \water\ and class~I \methanol\ masers through more energetic outflows.

\acknowledgments
We thank the anonymous referee for her/his constructive comments and suggestions that have helped to improve this paper. We are grateful to all staff members in Korean VLBI Network (KVN) who helped to operate the array and to correlate the data. The KVN is a facility operated by Korea Astronomy and Space Science Institute (KASI). We also thank Ray Furuya for providing the modified data of the maser and bolometric luminosities of their sample. We want to give another thank to James S. Urquhart for private discussions. 

\vspace{5mm}
\facilities{KVN 21 m} 
\software{GILDAS/CLASS (\citealt{pety2005}, GILDAS team)}



\bibliography{ms_uchii_maser_WK}



\begin{figure*}
	\centering
	\scalebox{0.19}[0.19]{\includegraphics{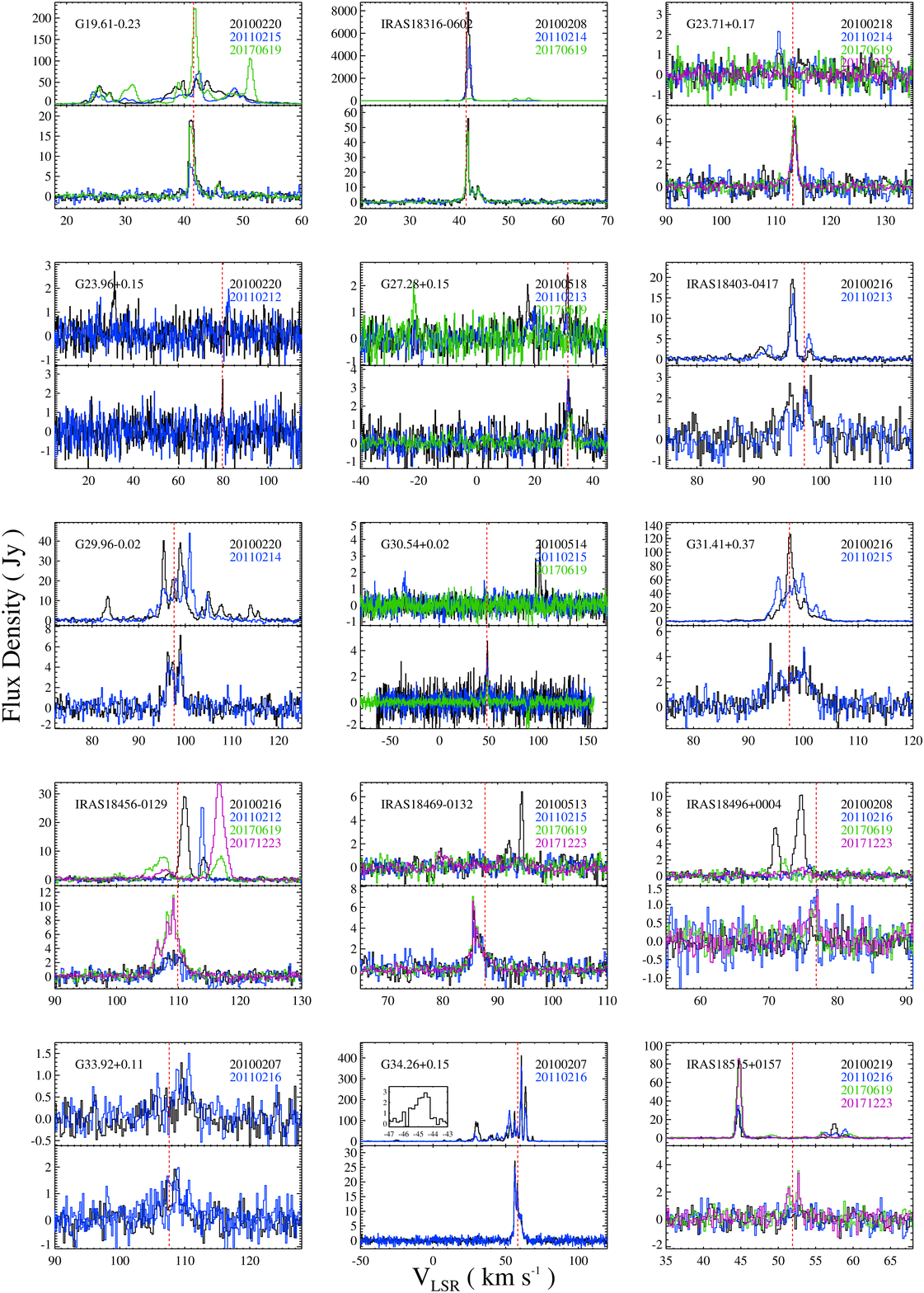}}

	Fig. 1.--- Continued.
\end{figure*}
\begin{figure*}
	\centering
	\scalebox{0.19}[0.19]{\includegraphics{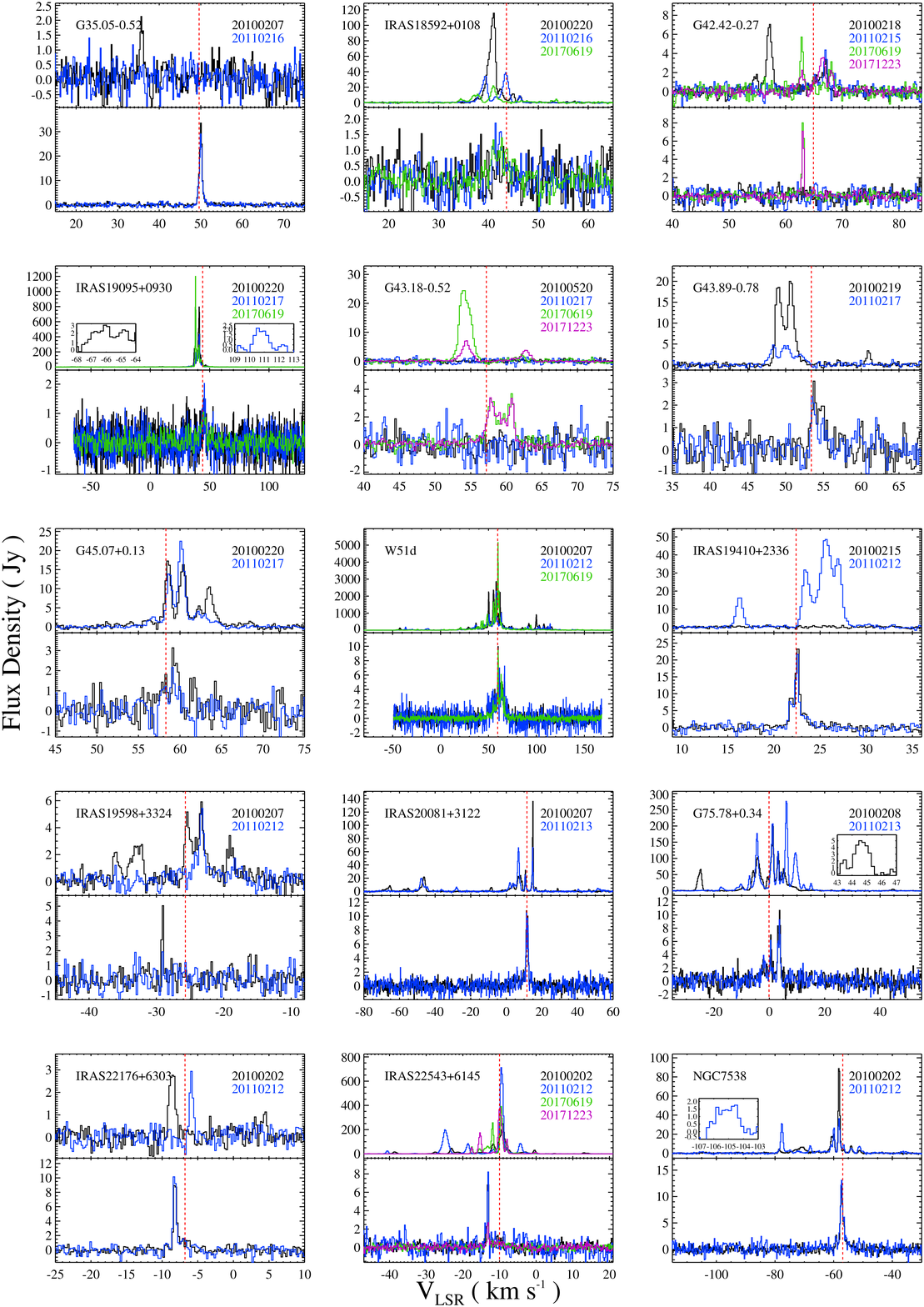}}

	Fig. 1.--- Continued.
\end{figure*}
\begin{figure*}
	\centering
	\scalebox{0.19}[0.19]{\includegraphics{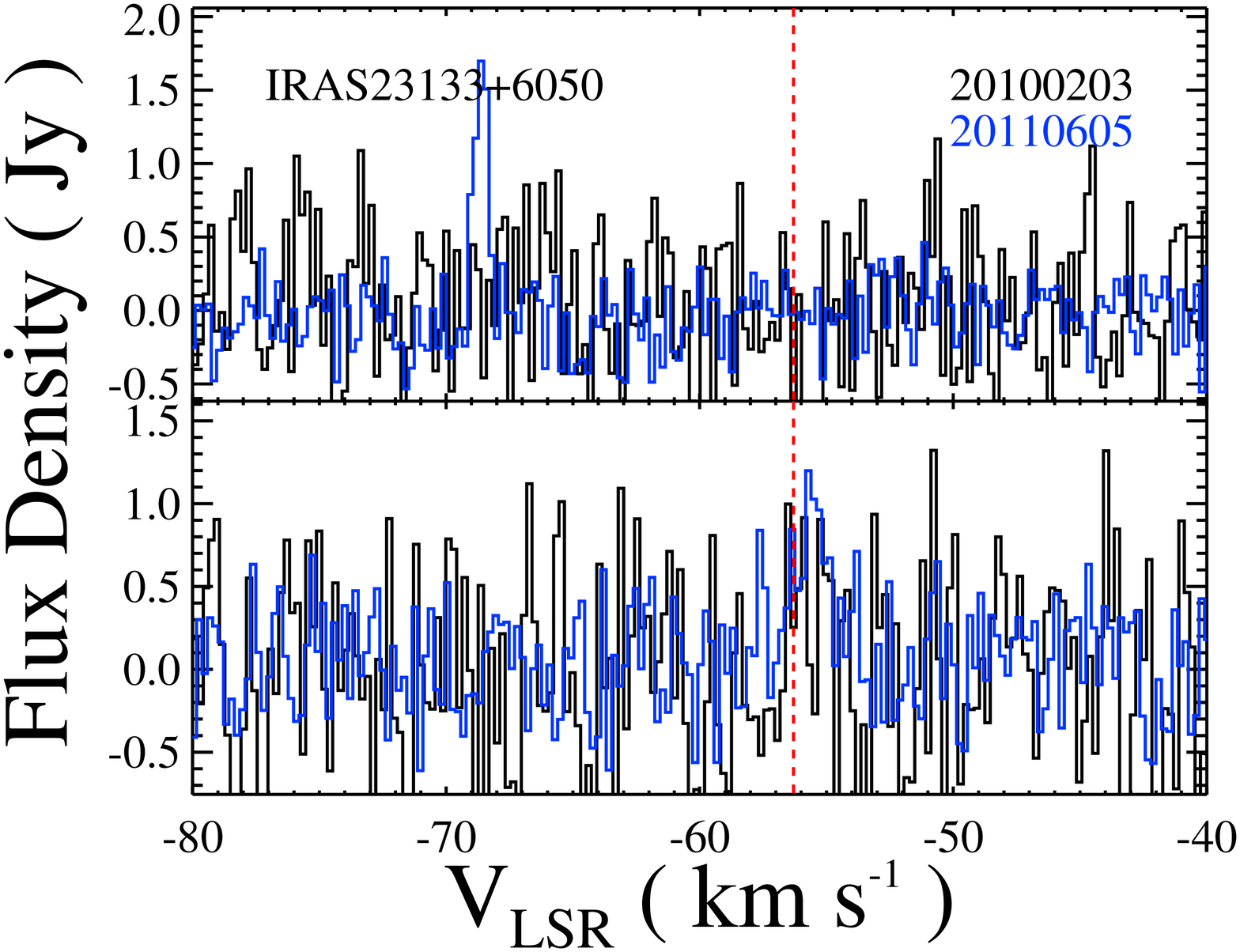}}
	Fig. 1.--- Continued.
\end{figure*}

\begin{figure*}
	\centering
    \scalebox{0.19}[0.19]{\includegraphics{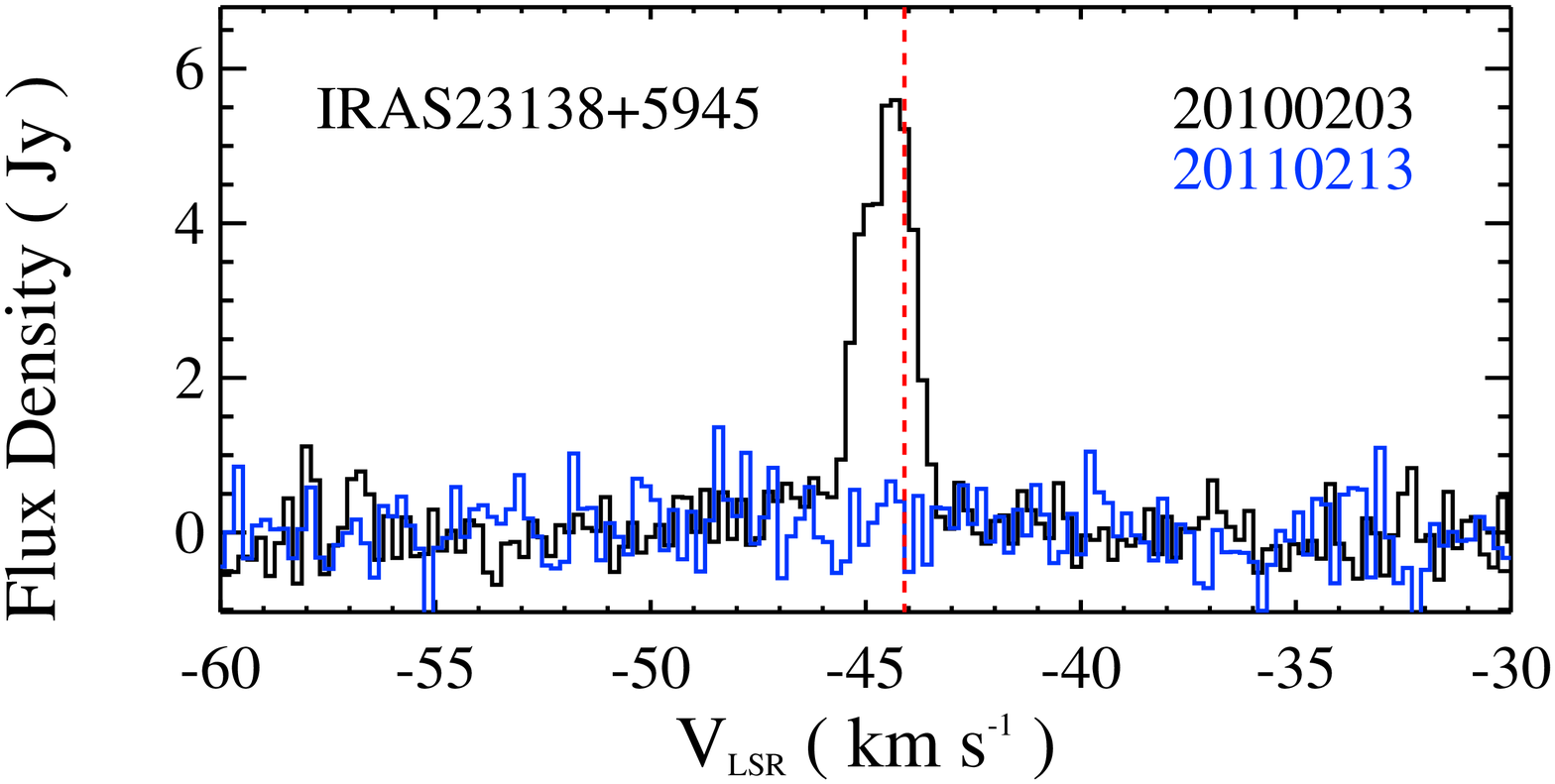}}
    Fig. 2 --- Continued.
\end{figure*}

\input table.inc

\clearpage
\appendix
\section{Map grids}

\begin{figure}[h!]
\centering
\scalebox{0.45}[0.45]{\includegraphics{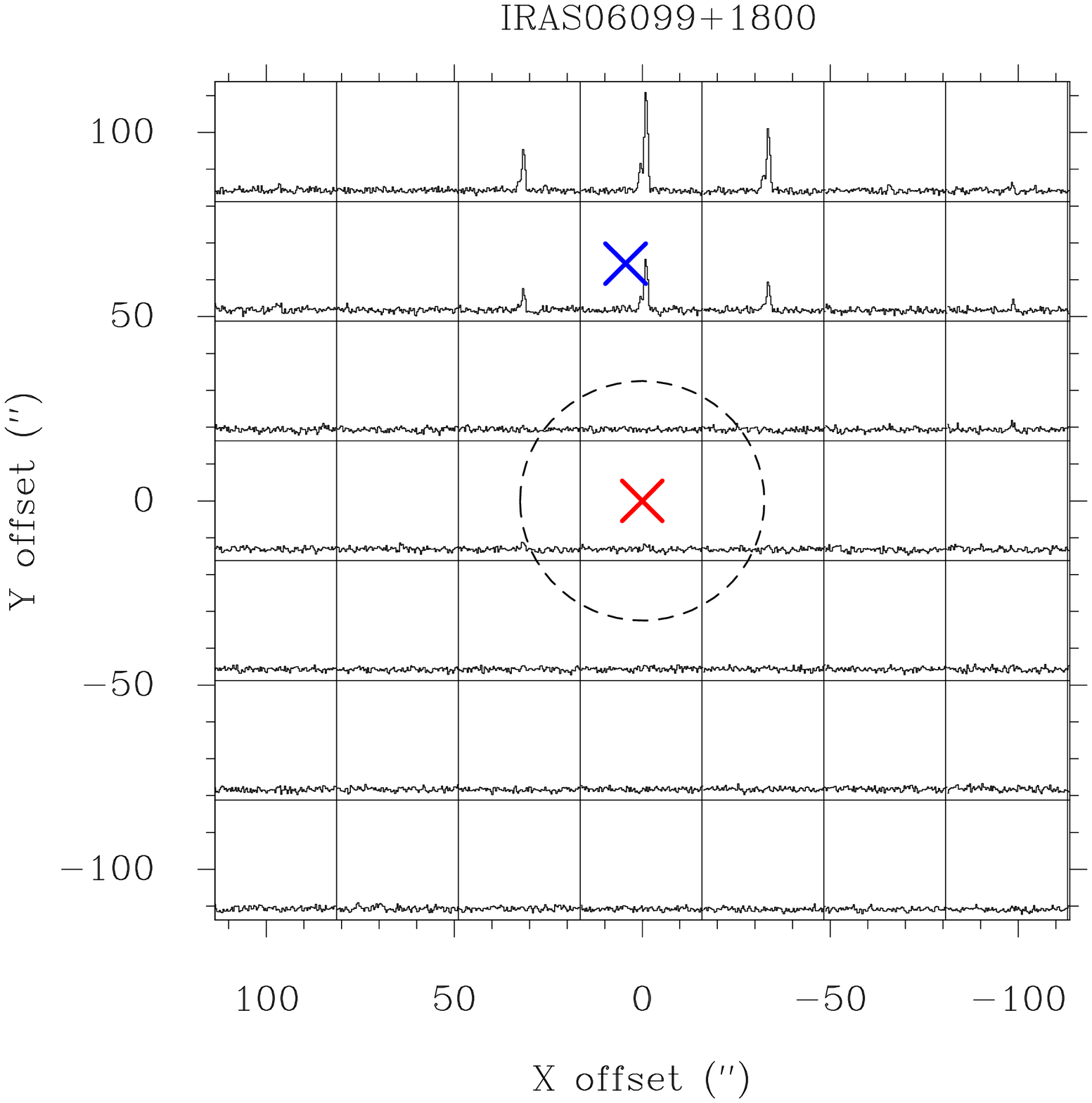}}
\scalebox{0.45}[0.45]{\includegraphics{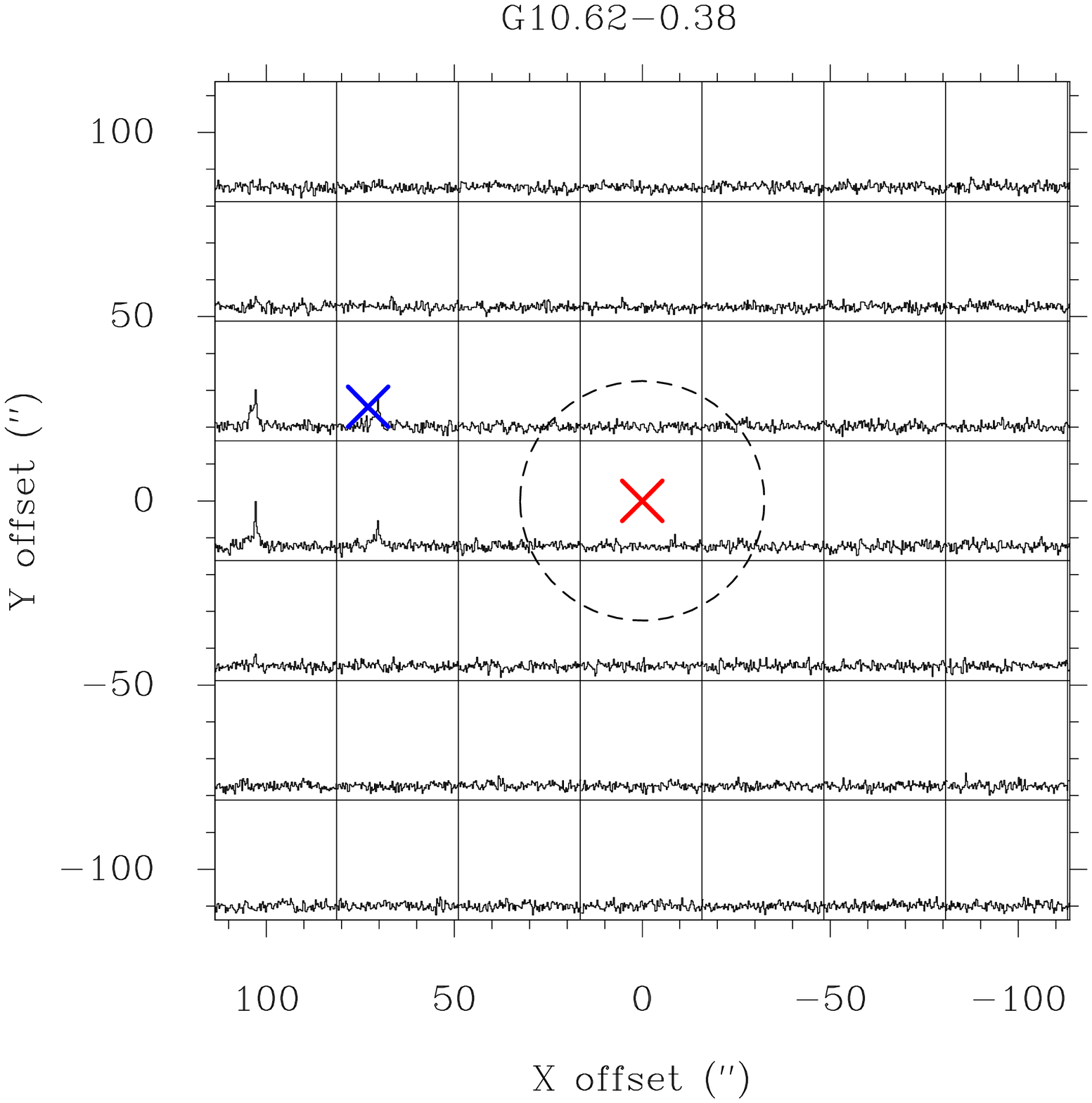}}
\caption{\label{appendix:mapping} Map grids of 44\,\ghz\ \methanol\ masers of IRAS06099$+$1800 and G10.62$-$0.38. The red crosses indicate observed positions at the 1st and 2nd epochs in 2010 and 2011. On the other hand, the blue ones are positions of radio continuum sources taken from the catalogs of WC89a and KCW94 and they were used at the 3rd and 4th epochs in 2017. The dashed-circles correspond to a FWHM beam size (65\arcsec) at 44\,\ghz.}
\end{figure}

\end{document}

%% file: table.inc

\clearpage

\begin{deluxetable*}{ l c c  c c c c c c c c c c c }
\centering
\tabletypesize{\tiny}
\setlength{\tabcolsep}{0.045in}
\setlength{\oddsidemargin}{-0.4in}
\tablewidth{0pt}
\tablecaption{\label{tb:source} Source Summary}
\tablehead{ 
\colhead{Source} &
\colhead{Ref.} &
\colhead{R.A. (2000)} &
\colhead{Dec. (2000)} &
\colhead{Offset} &
\colhead{Station} &
\colhead{Observing} &
\multicolumn{2}{c}{Maser detection}&
\colhead{$\sigma_{\rm H_{2}O}$} & 
\colhead{$\sigma_{\rm CH_{3}OH}$} &
\colhead{6.7\,GHz} &
\colhead{Other name}\\
\cline{8-9}
\colhead{Name} &
\colhead{} &
\colhead{($\emph{h m s}$)} &
\colhead{(\arcdeg \ \arcmin \ \arcsec)} & 
\colhead{\arcsec} &
\colhead{} &
\colhead{Date} & 
\colhead{H$_{2}$O} &
\colhead{CH$_{3}$OH} &
\colhead{(Jy)} &
\colhead{(Jy)} &
\colhead{Maser} &
\colhead{}\\
\colhead{(1)} & 
\colhead{(2)} &
\colhead{(3)} &
\colhead{(4)} &
\colhead{(5)} &
\colhead{(6)} &
\colhead{(7)} &
\colhead{(8)} &
\colhead{(9)} &
\colhead{(10)} &
\colhead{(11)} &
\colhead{(12)} &
\colhead{(13)}
}
\startdata
IRAS02232$+$6138& 2 & 02:27:01.0 & $+$61:52:14  &23&YS& 2010 Feb 06 & Y & Y & 0.38 & 0.52  && G133.947$+$1.064, W3OH \\
	        &   & 		 &                &&YS& 2011 Feb 12 & Y & Y & 1.15 & 0.37  &&\\
                &   & 02:27:03.8 & $+$61:52:25    &&YS& 2017 Jun 16 & Y & Y & 0.49 & 0.24  &Y$^{\ast}$&\\
                &   &            &                &&YS& 2017 Dec 23 & Y & Y & 0.35 & 0.20  && \\
IRAS02575$+$6017& 2 & 03:01:32.3 & $+$60:29:12  &15&YS& 2010 Feb 06 & N & Y$^{a}$ & 0.39 & 0.53  && G138.295$+$1.555, W5E \\
  	        &   & 		 & 		  &&YS& 2011 Feb 13 & N & N & 0.31 & 0.44  &&\\
                &   & 03:01:34.3 & $+$60:29:14    &&US& 2017 Jun 19 & N & N & 0.29 & 0.21  &&\\
                &   &            &                &&YS& 2017 Dec 23 & N & N  & 0.29 & 0.21 &&\\
\enddata
\tablecomments{ Source name in Column (1), original catalogs (1: WC89a, 2: KCW94) in Column (2), observed equatorial coordinates in Columns (3) and (4), \\ offsets between the observed coordinates in the 1st/2nd and the 3rd/4th epochs in Column (5), observation KVN station (YS: Yonsei, US: Ulsan, TN: Tamna) \\ in Column (6), observing date in Column (7), maser detection in Columns (8) and (9), 1{$\sigma$} rms noise levels in Columns (10) and (11), 6.7~\ghz\ Class~II \methanol\ \\ maser detection from other surveys (\citealt{xu2003,pandian2007,caswell2010,green2010,green2012,szymczak2012,hou2014};\\ \citealt{breen2015,hu2016}) within a circle of 65$''$ diameter in Column (12), other names in Column (13). \\ In Column (12), an asterisk ($\ast$) indicates 6.7\,\ghz\ detection found within a circle with a radius of 5$''$, and a dagger mark ($\dagger$) shows MMB 6.7\,\ghz\ sources. \\ In addition, a double dagger ($\ddagger$) means that there are MMB masers found within a radius of 32.5$''$ but not of 5$''$. }
\tablenotetext{a}{Newly detected objects from our survey.}
\tablenotetext{b}{Detected by the first sidelobe.}
\tablenotetext{c}{Non-detection due to a large offset between the coordinate used for observations in 2010 and 2011 and the corrected coordinate in 2017.}
\tablenotetext{d}{Detected by the first sidelobe in the first and second epochs, but detected by the main beam in the third and forth epochs} 
\end{deluxetable*}

\newpage
\begin{table*}
\centering \caption{\label{tb:detection}Detection Rates of \water\ and \methanol\ Masers}
\scriptsize
\begin{tabular}{ l l c c c c}
\hline \hline
Obs epoch & Source Group & & \# of Obs & \# of {\water} source (\%) &\# of {\methanol} source (\%) \\
\hline
2010 & Total    &	& 102 & 63 (62$\pm$9\%) & 41 (40$\pm$10\%)\\
     & WC89a    & 	& 51  & 35 (69$\pm$13\%) & 24 (47$\pm$14\%)\\      
     & KCW94    &	& 51  & 28 (55$\pm$14\%) & 17 (33$\pm$13\%)\\
\hline
2011 & Total    &	& 102 & 61 (60$\pm$10\%) & 37 (36$\pm$9\%)\\ 
     & WC89a    &	& 51  & 32 (63$\pm$13\%) & 21 (41$\pm$14\%)\\
     & KCW94    &	& 51  & 29 (57$\pm$14\%) & 16 (31$\pm$13\%)\\
\hline
2017 Jun & Total &  & 44  &    26 (59$\pm$15\%)  &  26 (59$\pm$15\%)   \\
         & WC89a     &  & 14  &    9 (64$\pm$25\%)  &   10  (71$\pm$24\%)  \\
         & KCW94     &  & 30  &    17 (57$\pm$18\%) &   16 (53$\pm$18\%)  \\
\hline
2017 Dec & Total  &  & 28  &   19 (68$\pm$17\%) &   15 (54$\pm$18\%)  \\
         &WC89a     &  & 6   &    4 (67$\pm$38\%)  &   5 (83$\pm$30\%)   \\
         &KCW94     &  & 22  &    15 (68$\pm$19\%) &   10 (45$\pm$21\%)  \\
\hline
All Obs & Total    &	& 103 & 74 (72$\pm$9\%) & 55 (53$\pm$10\%)\\
        & WC89a    &	& 52  & 38 (73$\pm$12\%) & 29 (56$\pm$13\%)\\
        & KCW94    &	& 51  & 36 (71$\pm$13\%) & 26 (51$\pm$14\%)\\
\hline
\end{tabular}
\tablecomments{The epoch is given in Column (1), subsample in Column (2), number of observed sources in Column (3), number of maser-detected sources in Columns (4)$-$(5).
The errors of the detection rates have been obtained by assuming binomial statistics.}
\end{table*}

\clearpage

\begin{deluxetable*}{ l c c c c r r r r r r r r r }
\tabletypesize{\scriptsize}
\setlength{\tabcolsep}{0.045in}
\setlength{\oddsidemargin}{-0.3in}
\tablewidth{0pt} 
\tablecaption{\label{tb:water_linepara} \water\ Maser Line Parameters of Detected 
Sources}
\tablehead{ 
\colhead{Source} &
\colhead{} &
\colhead{} &
\colhead{Observing} &
\colhead{~~~~} &
\colhead{$S_{\rm p}$} & 
\colhead{} &
\colhead{$V_{\rm p}$} &
\colhead{} &
\colhead{$V_{\rm low}$} &
\colhead{} &
\colhead{$V_{\rm high}$} &
\colhead{} &
\colhead{$\int {S_{\rm \nu}~dv}$} \\
\colhead{Name} &
\colhead{} &
\colhead{} &
\colhead{Date} &
\colhead{~~~~} &
\colhead{(Jy)} &
\colhead{} &
\colhead{(\kms)} &
\colhead{} &
\colhead{(\kms)} &
\colhead{} &
\colhead{(\kms)} &
\colhead{} &
\colhead{(Jy~\kms)} \\
\colhead{(1)} &
\colhead{} &
\colhead{} &
\colhead{(2)} &
\colhead{~~~~} &
\colhead{(3)} &
\colhead{} &
\colhead{(4)} &
\colhead{} &
\colhead{(5)} &
\colhead{} &
\colhead{(6)} &
\colhead{} &
\colhead{(7)}
} 

\startdata
IRAS02232$+$6138 &&& 2010 Feb 06 &&4466.4&&$-$48.3&& $-$72.5&& $-$37.8&& 8496.4 \\
	             &&& 2011 Feb 12 &&3349.4&&$-$48.4&& $-$99.3&& $-$44.6&& 5570.8 \\
                 &&& 2017 Jun 16 && 670.9&&$-$47.6&& $-$84.8&& $-$33.4&& 2496.7 \\
                 &&& 2017 Dec 23 && 796.8&&$-$54.8&& $-$93.2&& $-$32.5&& 3169.4 \\
IRAS03035$+$5819 &&& 2011 Feb 13 &&   2.4&&$-$32.1&& $-$32.4&& $-$25.3&& 3.5 \\
			     &&& 2017 Jun 20 &&   3.1&&$-$37.6&& $-$37.6&& $-$37.2&& 2.4 \\
                 &&& 2017 Dec 23 &&   9.1&&$-$37.5&& $-$37.9&& $-$36.9&& 6.3 \\
IRAS05391$-$0152 &&& 2017 Dec 23 &&   2.6&&   18.8&&    17.8&&    19.1&& 2.5 \\
\enddata
\tablecomments{Source name in Column (1), observing date in Column (2),
peak flux density, peak velocity, lowest velocity, highest velocity,
and integrated flux density in Columns (3)$-$(7).}
\end{deluxetable*}

\clearpage
\begin{deluxetable*}{l c c c c r r r r r r r r r}
\tabletypesize{\scriptsize}
\setlength{\tabcolsep}{0.045in}
\setlength{\oddsidemargin}{-0.3in}
\tablewidth{0pt} 
\tablecaption{\label{tb:methanol_linepara} \methanol\ Maser Line Parameters of Detected Sources}
\tablehead{
\colhead{Source} &
\colhead{} &
\colhead{} &
\colhead{Observing} &
\colhead{~~~~} &
\colhead{$S_{\rm p}$} &
\colhead{} &
\colhead{$V_{\rm p}$} &
\colhead{} &
\colhead{$V_{\rm low}$} &
\colhead{} &
\colhead{$V_{\rm high}$} &
\colhead{} &
\colhead{$\int {S_{\rm \nu}~dv}$} \\
\colhead{Name} &
\colhead{} &
\colhead{} &
\colhead{Date} &
\colhead{~~~~} &
\colhead{(Jy)} &
\colhead{} &
\colhead{(\kms)} &
\colhead{} &
\colhead{(\kms)} &
\colhead{} &
\colhead{(\kms)} &
\colhead{} &
\colhead{(Jy\,\kms)} \\
\colhead{(1)} &
\colhead{} &
\colhead{} &
\colhead{(2)} &
\colhead{~~~~} &
\colhead{(3)} &
\colhead{} &
\colhead{(4)} &
\colhead{} &
\colhead{(5)} &
\colhead{} &
\colhead{(6)} &
\colhead{} &
\colhead{(7)}
}

\startdata
IRAS02232$+$6138&&& 2010 Feb 06 && 5.2&& $-$46.4 && $-$47.8 && $-$45.0 && 9.7  \\
	            &&& 2011 Feb 12 && 3.2&& $-$46.2 && $-$48.0 && $-$45.3 && 6.9  \\
                &&& 2017 Jun 16 && 1.9&& $-$46.5 && $-$47.7 && $-$45.4 && 4.7 \\ 
                &&& 2017 Dec 23 && 2.5&& $-$46.5 && $-$49.6 && $-$45.4 && 6.3 \\
IRAS02575$+$6017&&& 2010 Feb 06 && 1.9&& $-$36.9 && $-$37.1 && $-$36.8 && 1.5  \\
IRAS05391$-$0152&&& 2017 Jun 17 && 1.0&& 10.6 && 9.1 && 10.6 && 1.3 \\
				&&& 2017 Dec 23 && 0.9&& 10.6 && 9.5 && 10.6 &&1.3 \\
MONR2           &&& 2017 Jun 20 && 0.9&& 10.7 && 10.7 && 10.7 && 1.8 \\
\enddata
\tablecomments{Source name in Column (1), observing date in Column (2), peak flux density, peak velocity, the lowest velocity, the highest velocity, and integrated flux density in Columns (3)$-$(7).}
\end{deluxetable*}

\clearpage

\begin{deluxetable*}{  l c c r r r r c c }
\tabletypesize{\scriptsize}
\centering
\setlength{\oddsidemargin}{-0.4in}
\tablewidth{0pt} 
\tablecaption{\label{tb:high-velo} \water\ Maser Sources with High-velocity ($>$\,30\,\kms) Lines}
\tablehead{ 
\colhead{Source} &
\colhead{} &
\colhead{Observing} &
\colhead{$V_{\rm sys}$} &
\colhead{$V_{\rm p}$} &
\colhead{$V_{\rm blue}$} &
\colhead{$V_{\rm red}$} &
\colhead{} &
\colhead{} \\
\colhead{Name} &
\colhead{} &
\colhead{Date} &
\colhead{(\kms)} &
\colhead{(\kms)} &
\colhead{(\kms)} &
\colhead{(\kms)} &
\colhead{} &
\colhead{Notes} \\
\colhead{(1)} &
\colhead{} &
\colhead{(2)} &
\colhead{(3)} &
\colhead{(4)} &
\colhead{(5)} &
\colhead{(6)} &
\colhead{} &
\colhead{(7)} 
}
\startdata
IRAS02232$+$6138 & & 2010 Feb 06 & $-46.4$ & $-48.3$ & $-71.3$ & $-38.8$ &	& blue-shifted HV  \\
		 & & 2011 Feb 12 &         & $-48.4$ & $-98.5$ & $-45.7$ &	& blue-shifted EHV \\
                 & & 2017 Jun 16 &         & $-$47.6 & $-$84.3 & $-$35.2 &      & blue-shifted HV   \\
                 & & 2017 Dec 23 &         & $-$54.8 & $-$92.1 & $-$33.4 &      & blue-shifted HV   \\
IRAS06061$+$2151 & & 2010 Feb 07 & $-0.7$  & $-7.4$  & $-12.7$ & 37.0    &	& red-shifted HV \\
		 & & 2011 Feb 13 &	   & $-7.5$  & $-13.4$ & 15.8    &	& none \\
                 & & 2017 Jun 20 &         & $-$5.9  & $-$19.5 & 26.9    &      & none               \\
                 & & 2017 Dec 23 &         & $-$7.2  & $-$18.1 & 22.6    &      & none               \\
\enddata
\tablecomments{Source name in Column (1), observing date in Column (2), systemic velocity in Column (3), peak velocity in Column (4), peak velocities of the most blue and redshifted lines in Columns (5){$-$}(6), note in Column (7).}
\end{deluxetable*}

\begin{deluxetable*}{ l c c c c c c c c c c c c c c c }
\tabletypesize{\scriptsize}
\setlength{\tabcolsep}{0.040in}
\tablewidth{0pt} \tablecaption{\label{tb:properties} Properties of Sources}
\tablehead{ 
\colhead{Source} &
\colhead{} &
\colhead{$V{_{\rm sys}}^{a}$} &
\colhead{Dist.$^{b}$} &
\colhead{} &
\colhead{${L_{\rm bol}}^{c}$} &
\colhead{} &
\colhead{${S_{\rm 2cm}}^{d}$} &
\colhead{} &
\colhead{${S_{\rm 6cm}}^{e}$} & 
\colhead{} &
\colhead{${S_{\rm 850{\mu}m}}^{f}$} &
\colhead{} &
\colhead{${L\rm{_{H_{2}O}}}$} & 
\colhead{} &
\colhead{$L\rm{_{CH_{3}OH}}$} \\
\colhead{Name} & 
\colhead{} &
\colhead{(\kms)} &
\colhead{(kpc)} &
\colhead{} &
\colhead{(L$_{\odot}$)} &
\colhead{} &
\colhead{(mJy)} &
\colhead{} & 
\colhead{(mJy)} &
\colhead{} & 
\colhead{(Jy)} &
\colhead{} & 
\colhead{(L$_{\odot}$)} &
\colhead{} &
\colhead{(L$_{\odot}$)}\\
\colhead{(1)} & 
\colhead{} &
\colhead{(2)} &
\colhead{(3)} &
\colhead{} &
\colhead{(4)} &
\colhead{} &
\colhead{(5)} &
\colhead{} &
\colhead{(6)} &
\colhead{} & 
\colhead{(7)} & 
\colhead{} &
\colhead{(8)} &  
\colhead{} &
\colhead{(9)}
}
\startdata
IRAS02232$+$6138 &	& $-46.4$ & 3.0 &	& 2.06 $\times 10^{5}$ &  & 2533.9   &  & $\cdots$ &  & 62.1     &  & [1.77, 1.16, 0.52, 0.66] $\times 10^{-3}$&	& [4.00, 2.85,1.95, 2.62] $\times 10^{-6}$ \\
IRAS02575$+$6017 &	& $-38.1$ & 3.8 &	& 3.70 $\times 10^{4}$ &  & $\cdots$ &  & 3.2      &  & 19.6     &  & $\cdots$  &	& [9.86, $\cdots$, $\cdots$, $\cdots$] $\times 10^{-7}$ \\
IRAS03035$+$5819 &	& $-40.0$ & 4.2 &	& 6.60 $\times 10^{4}$ &  & 18.5     &  & $\cdots$ &  & $\cdots$ &  & [$\cdots$, 1.41, 0.99, 2.56] $\times 10^{-6}$ &	& $\cdots$  \\
IRAS05391$-$0152 &  & 10.0    & 0.7 &   & $\cdots$             &  & $\cdots$ &  & $\cdots$ &  & $\cdots$ &  & [$\cdots$,$\cdots$,$\cdots$,2.81]$\times 10^{-8}$ &   & [$\cdots$,$\cdots$,3.04, 2.88]$\times 10^{-8}$\\
IRAS05393$-$0156 &  & 10.0    & 0.5 &   & 1.30 $\times 10^{4}$ &  & 19.5     &  & $\cdots$ &  & $\cdots$ &  & [$\cdots$,$\cdots$,3.40, 3.32]$\times 10^{-8}$ &  & $\cdots$ \\
MON R2           &  & 10.8    & 0.8 &	& 2.91 $\times 10^{4}$ &  & 333.6    &  & 3129.9   &  & $\cdots$ &  & [--, --, 0.79, 1.00] $\times 10^{-5}$ &	& [$\cdots$,$\cdots$,5.16, 3.87]$\times 10^{-8}$  \\
IRAS06058$+$2138 &	& 3.1     & 2.2 &	& $\cdots$             &  & $\cdots$ &  & 2.0      &  & 24.4     &  & [1.43, 1.00, 0.03, 0.10] $\times 10^{-5}$ &	& $\cdots$ \\
IRAS06061$+$2151 &	& $-0.7$  & 4.1 &	& $\cdots$             &  & $\cdots$ &  & $\cdots$ &  & 8.4      &  & [2.39, 3.29,4.83,4.98] $\times 10^{-4}$ &	& $\cdots$  \\
\enddata
\tablecomments{Systemic velocity measured from molecular lines surveys presents in Column (2), kinematic distance is in Column (3), bolometric luminosity \\ central object is in Column (4), radio continuum flux densities at 2 and 6\,cm are listed in Columns (5)-(6), and submillimeter continuum \\ flux density at 850\,${\mu}m$ is in Column (7).}
\tablenotetext{a}{~References for column (2) are \cite{churchwell1990,wouterloot1993,anglada1996,bronfman1996};\\ \cite{shepherd1996,purcell2006}.}
\tablenotetext{b}{~References for column (3) are \cite{wood1989a,palagi1993,hofner1996,kurtz1994};\\ \cite{caswell1995,harju1998,beuther2003,watson2003,xu2003,herbst1976}.}
\tablenotetext{c}{~References for column (4) are \cite{wood1989a,kurtz1994}, and a reference for MonR2 is \cite{connelley2008}.}
\tablenotetext{d}{~References for column (5) are \cite{wood1989a,kurtz1994}.}
\tablenotetext{e}{~References for column (6) are \cite{simon1981,snell1986,wood1989a,carpenter1990}.}
\tablenotetext{f}{~References for column (7) are \cite{thompson2006}.}
\end{deluxetable*}